\documentclass{article}

\usepackage{graphicx}
\usepackage{float}
\usepackage{amsmath,amssymb}
\usepackage{color}
\usepackage{caption}
\usepackage{wrapfig}
\usepackage{cite}

\begin{document}

\title{\includegraphics[width=0.35\textwidth]{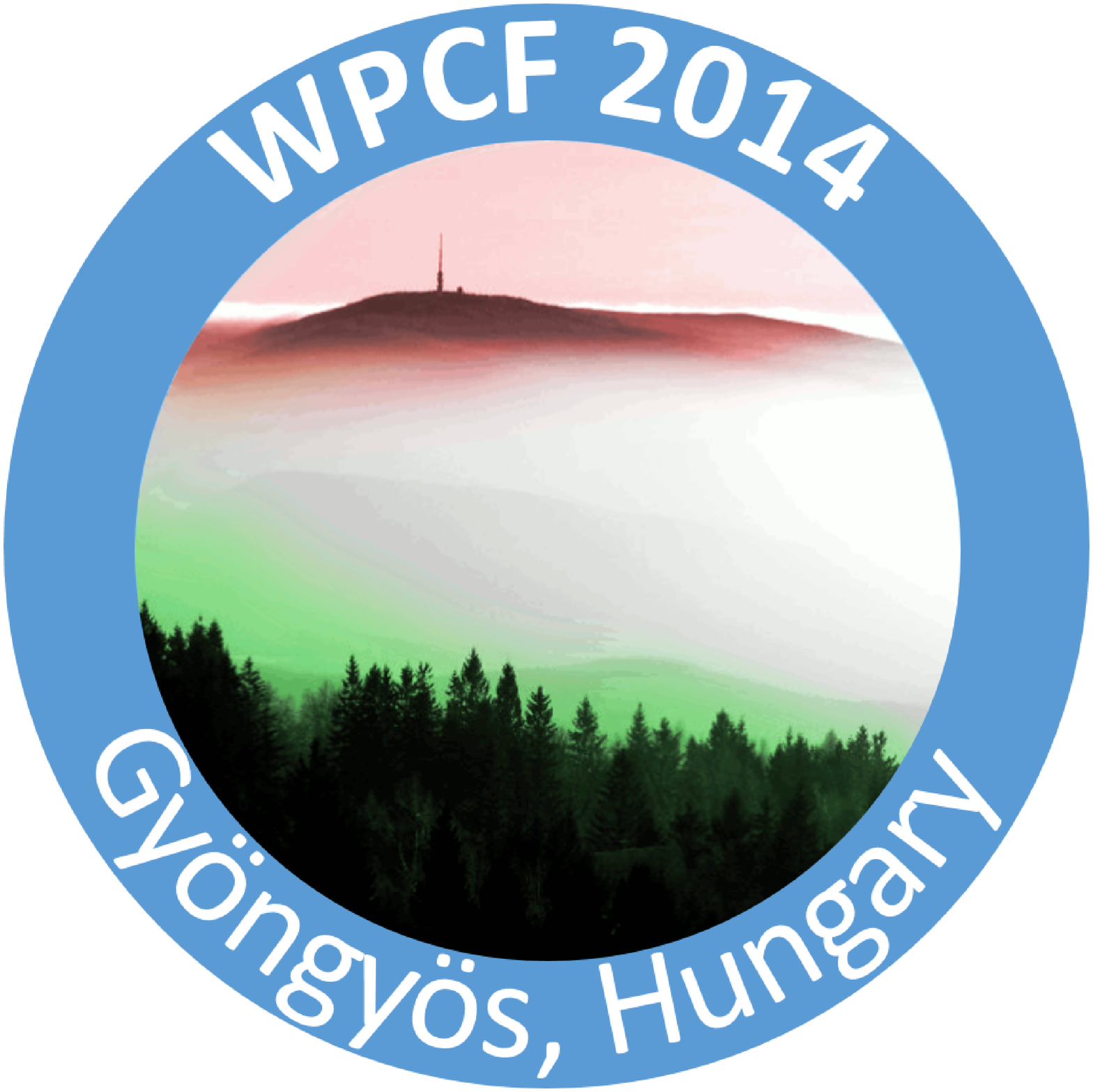}\\[1cm]
Excitation function of elastic $pp$ scattering\\from a unitarily extended Bialas-Bzdak model}

\author{F. Nemes$^{1,2}$\\
{\small $^1$ CERN, CH-1211 Geneva 23, Switzerland\vspace{6mm}}\\
T. Cs\"org\H{o}$^{2,3}$\\
{\small $^2$ Wigner Research Centre for Physics, Hungarian Academy of Sciences}\\ {\small H-1525 Budapest 114, P.O.Box 49, Hungary}\\
{\small $^3$KRF, H-3200 Gy\"ongy\"os, M\'atrai \'ut 36, Hungary}\vspace{6mm}\\
M. Csan\'ad$^{4}$\\
{\small $^{4}$ E\"otv\"os University, Department of Atomic Physics}\\{\small H-1117 Budapest, P\'azm\'any P\'eter s., 1/A Hungary}}

\maketitle

\begin{abstract}
	The Bialas-Bzdak model of elastic proton-proton scattering assumes a purely
	imaginary forward scattering amplitude, which consequently vanishes at the diffractive minima.
	We extended the model to arbitrarily large real parts in a way that constraints from
	unitarity are satisfied.  The resulting model is able to describe elastic
	$pp$ scattering not only at the lower ISR energies but also at $\sqrt{s}=$7 TeV
	in a statistically acceptable manner, both in the diffractive cone and in the
	region of the first diffractive minimum.  The total cross-section as well as the
	differential cross-section of elastic proton-proton
	scattering is predicted for the future LHC energies of $\sqrt{s}=$8, 13, 14, 15 TeV and also to 28 TeV.  A non-trivial, significantly
	non-exponential feature of the differential cross-section of elastic proton-proton
	scattering is analyzed and the excitation function of the non-exponential 
	behavior is predicted.  The excitation function of the shadow profiles 
	is discussed and related to saturation at small impact parameters. 

\end{abstract}

\section{Introduction}

In a pair of recent papers the Bialas-Bzdak model~\cite{Bialas:2006qf} (BB) of small angle elastic proton-proton ($pp$)
scattering at high energies was studied at 7 TeV LHC energy~\cite{Nemes:2012cp,CsorgO:2013kua}. In those papers a terse
overview is reported about the field of elastic scattering at high energies. Here we would like to highlight only some
recent works which influenced us.

In this manuscript the BB model is extrapolated to future LHC energies. Our method to include the energy evolution of
the parameters is somewhat similar to the so-called ``geometric scaling'' discussed in Ref.~\cite{Ferreira:2014gda} and also
in Ref.~\cite{Kohara:2014waa}.

Using 2012 data the TOTEM experiment recently made an important experimental observation at $\sqrt{s}=8$~TeV: the $pp$ elastic differential cross-section shows a deviation from the most
simple non-exponential behavior at low-$|t|$,~\cite{Simone:WPCF2014} where $t$ is the squared four-momentum transfer of the $pp$ scattering process.
This feature of the $\sqrt{s}=8$~TeV (preliminary) TOTEM dataset, was related to $t$-channel unitarity of the forward scattering amplitude (FSA) in Ref.~\cite{Jenkovszky:2014yea}, a concept that
we also focus on, using and generalizing in a unitary manner  the quark-diquark model
of Bialas and Bzdak for the determination of the shape of the FSA of elastic $pp$ scattering.

In its original form, the BB model~\cite{Bialas:2006qf} assumes that the
real part of the FSA is negligible,
correspondingly, the FSA vanishes at the diffractive minima. At the ISR energies of $\sqrt{s}=$23.5$-$62.5 GeV, that were first analyzed
in the inspiring paper of  Bialas and Bzdak~\cite{Bialas:2006qf}, this assumption is indeed reasonable,
as confirmed in Ref.~\cite{Nemes:2012cp}. At these ISR energies, only very few data points were available in the dip region around the
first diffractive minimum of elastic $pp$ scattering, which were then left out from
the BB model fits of Ref.~\cite{Nemes:2012cp} to achieve a quality description
of the remaining data points.
However, in recent years, TOTEM data~\cite{Antchev:2011vs} explored the dip region at the LHC energy of 7 TeV in
great details,  at several different values of the squared four-momentum transfer $t$.
Ref.~\cite{Nemes:2012cp} demonstrated, that the original BB model cannot
describe this dip region, not without at least a small real part that has to be
added to its FSA in a reasonable way.

Subsequently, the BB model has been generalized in Ref.~\cite{CsorgO:2013kua} 
by allowing for a perturbatively small real part of the FSA, 
which improved the agreement of the model with TOTEM data on elastic $pp$
scattering at the LHC energy of $\sqrt{s} = 7$ TeV. 
It was expected that the main reason for the appearance
of this real part is that certain rare elastic scattering of the constituents
of the protons may be non-collinear thus may lead to inelastic events even if
the elementary interactions are elastic.  
The corresponding phenomenological generalization of the Bialas-Bzdak model~\cite{CsorgO:2013kua} was indeed based on the assumption that the
real part of the FSA is small, and can be handled
perturbatively.  The resulting $\alpha$-generalized Bialas-Bzdak 
($\alpha$BB) model was compared to ISR data in
Ref.  ~\cite{CsorgO:2013kua}, and it was demonstrated that a small, 
of the order of 1 \% real part of the FSA indeed 
results in excellent fit qualities and a
statistically acceptable description of the data in the region of the
diffractive minimum or dip.  However, at the LHC energy of 7 TeV, the
same $\alpha$BB model does not result in a satisfactory, statistically acceptable fit
quality, although the visual quality of the fitted curves improve 
significantly as compared to that of the original BB model~\cite{CsorgO:2013kua}. 
	
	These results indicate that at the LHC energies the real part of the
FSA may reach significant values where unitarity constraints may already play an important
role. The unitarity of the $S$-matrix provides also the basis for the optical 
theorem, which in turn provides a method to determine the total cross-section from 
an extrapolation of the elastic scattering measurements to the $t = 0$ point.

	In the $\alpha$BB model of Ref.~\cite{CsorgO:2013kua}, 
unitarity constraints were not explicitly considered: as the original BB model
with zero real parts obeyed unitarity, adding a small real part may possibly
resulted only in small violations of unitarity and the optical theorem. 
However, when the model was fitted to the 7 TeV TOTEM data in the dip region in
Ref.~\cite{CsorgO:2013kua}, the extrapolation to the point of $t = 0$
and the related value of the total cross-section underestimated the measured
total cross-section by about 40\%, suggesting, that perhaps the real part of
the FSA may be large, and unitarity relations should be explicitly considered. 

	These indications motivate the present manuscript, where the Bialas-Bzdak model
is further generalized to arbitrarily large real parts of the FSA, fully taking into account  unitarity constraints.  The resulting model is referred
to as the real extended Bialas-Bzdak (ReBB) model. 

The structure of the manuscript is as follows: in Section~\ref{sec:unitarity}, 
the general form of the forward scattering amplitude is re-derived for the case of
a non-vanishing real part starting from $S$-matrix unitarity. Then this result is applied to the extension of the BB model to
a non-vanishing and possibly large real part of the FSA.  

	In Section~\ref{sec:fitdescription}, the resulting ReBB model is fitted to
TOTEM data on elastic $pp$ scattering at $\sqrt{s} = $ 7 TeV, both in the
diffractive cone~\cite{Antchev:2013gaa,Antchev:2013iaa} and in the dip
region~\cite{Antchev:2011vs}, separately.

Based on these fits and comparisons of the ReBB model to $\sqrt{s}=$ 7 TeV data, in Section~\ref{sec:shadow_profile} the shadow profile function A(b) is evaluated. This function characterizes
the probability of inelastic $pp$ scattering at a given impact parameter $b$, and is compared to the shadow
profile functions of elastic $pp$ collisions at lower, ISR energies. Section \ref{sec:non_exponential} is
devoted to study the structure of the differential cross-section $d\sigma/dt$ at low-$|t|$ values and also to compare it with a purely exponential behavior.

	In Section~\ref{sec:excitation}, the excitation function of 
the fit parameters is investigated and their evolution with $\sqrt{s}$ 
is obtained based on a geometrical picture. The model parameters are extrapolated 
to the expected future LHC energies of 8, 13, 14 and 15 TeV, 
as well as for 28 TeV, that is not foreseen to be available 
at man-made accelerators in the  near future, but may be relevant 
for the investigation of cosmic ray events. The excitation functions of the shadow profile functions~$A(b)$
are also discussed. Finally we summarize and conclude.

\section{The real extended Bialas-Bzdak model}
\label{sec:unitarity} 

Although the original form of the Bialas-Bzdak model neglects the real part of
the FSA in high energy elastic $pp$ scattering, the
model is based on Glauber scattering theory and obeys unitarity constraints. 

The phenomenological generalization of the Bialas-Bzdak 
model~\cite{CsorgO:2013kua} is based on the assumption, that the
real part of the FSA is small, and can be handled
perturbatively, so unitarity constraints are not violated strongly.
However, it turned out that the addition of a small real part does not lead to a
statistically acceptable description of TOTEM data on elastic $pp$ collisions at~$\sqrt{s} = $ 7 TeV. 
In this manuscript, we consider the case, when the real part of the
FSA is not perturbatively small. 
We restart from $S$-matrix unitarity,
and consider how the BB model can be extended
to significant, real values of the FSA while satisfying the constraints of 
unitarity.

\subsection{S-matrix unitarity in the context of elastic proton-proton scattering}
\label{unitarity}
In this subsection some of the basic equations of quantum scattering theory are recapitulated.
The scattering or $S$ matrix describes how 
a physical system changes in a scattering process. 
The unitarity of the $S$ matrix ensures that the sum of the 
probabilities of all possible outcomes of the scattering process is one.

	The unitarity of the scattering matrix~$S$ is expressed by the equation
	\begin{equation}
	 	SS^{\dagger}=I\,,
		\label{S_matrix_unitarity}
	\end{equation}
	where $I$ is the identity matrix. The decomposition $S=I + iT$, where $T$ is the transition matrix, leads the unitarity relation Eq.~(\ref{S_matrix_unitarity}) to
	\begin{equation}
	 	T - T^{\dagger}=iTT^{\dagger}\,,
		\label{T_matrix_unitarity}
	\end{equation}
	which can be rewritten in the impact parameter $b$ representation as
	\begin{equation}
	 	2\,\text{Im}\,t_{el}(s,b)=|t_{el}(s,b)|^{2} + \tilde\sigma_{inel}(s,b)\,,
		\label{master_equation}
	\end{equation}
	where $s$ is the squared total center-of-mass energy.

	The functions $\tilde\sigma_{inel}(s,b)=d^{2}\sigma_{inel}/d^{2}b$ and $|t_{el}(s,b)|^{2}=d^{2}\sigma_{el}/d^{2}b$ are the inelastic and elastic scattering probabilities
	per unit area, respectively. The elastic amplitude $t_{el}(s,b)$ is defined in the impact parameter space and corresponds to the $\ell$th partial wave amplitude $T_{\ell}(s)$ through the relation $\ell+1/2\leftrightarrow b\sqrt{s}/2$, which
	is valid in the high energy limit, $\sqrt{s}\rightarrow\infty$.

	The unitarity relation~(\ref{master_equation}) is a second order polynomial equation in terms of the (complex) elastic amplitude $t_{el}(s,b)$. If
	one introduces the opacity or eikonal function~\cite{Glauber_lectures, Levin:1998pk, Khoze:2014aca, Ryskin:2012az, Ryskin:2009qf, Martin:2012nm}
	\begin{align}
 		\label{uBB_ansatz}
		t_{el}(s,b)&=i\left[1-e^{-\Omega(s,b)}\right]\,,
	\end{align}
	$\tilde\sigma_{inel}$ can be expressed as
	\begin{align}
 		\label{uBB_ansatz}
	 	\tilde\sigma_{inel}(s,b)&=1-e^{-2\,\text{Re}\,\Omega(s,b)}\notag\,.
	\end{align}
	The formula for $t_{el}$ is the so called eikonal form. From Eq.~(\ref{uBB_ansatz}) the real part of the opacity function $\Omega(s,b)$ can be expressed as
	\begin{equation}
	 	\text{Re}\,\Omega(s,b)=-\frac{1}{2}\ln\left[1-\tilde\sigma_{inel}(s,b)\right]\,.
		\label{connection_omega_sigma}
	\end{equation}
	In the original BB model it is assumed that the real part of $t_{el}$ vanishes. In this case Eqs.~(\ref{uBB_ansatz}) and~(\ref{connection_omega_sigma}) implies that
	\begin{align}
	 	t_{el}(s,b)=i\left[1-\sqrt{1-\tilde\sigma_{inel}(s,b)}\right]\,.
		\label{original_BB_model}
	\end{align}
	If the imaginary part $\text{Im } \Omega$ is taken into account in Eq.~(\ref{uBB_ansatz}) the result is
	\begin{align}
	 	&t_{el}(s,b)=i\left[1-e^{-i\,\text{Im}\,\Omega(s,b)}\sqrt{1- \tilde\sigma_{inel}(s,b)}\right]\,,
		\label{unitarized_BB_model}
	\end{align}
	where the concrete parametrization of $\text{Im}\,\Omega(s,b)$ is discussed later.

	To compare the model with data the amplitude~Eq.~(\ref{unitarized_BB_model}) has to be transformed into momentum space
	\begin{align}
		T(s,\Delta)&= \int\limits^{+\infty}_{-\infty}\int\limits^{+\infty}_{-\infty}{e^{i{\vec \Delta} \cdot {\vec b}}{t_{el}(s,b)\text{\rm d}^2b}}\\
		&= 2\pi i\int\limits_0^{\infty}{J_{0}\left(\Delta\cdot b\right)\left[1-e^{-\Omega\left(s,b\right)}\right]b\, {\rm d}b}\,,
		\label{elastic_amplitude}
	\end{align}
	where $b=|{\vec b}|$, $\Delta=|{\vec\Delta}|$ is the transverse momentum and $J_{0}$ is the zero order Bessel-function of the first kind. In the high energy limit, $\sqrt{s}\rightarrow\infty$, $\Delta(t) \simeq \sqrt{-t}$  where $t$ is the squared four-momentum
	transfer. Consequently the elastic differential cross-section can be evaluated as 
	\begin{equation}
		\frac{{\rm d}\sigma}{{\rm d} t}=\frac{1}{4\pi}\left|T\left(s,\Delta\right)\right|^2\,.
		\label{differential_cross_section}
	\end{equation}

	According to the optical theorem the total elastic cross-section is
	\begin{align}
		\sigma_{tot}=2 \left.T(s,\Delta)\right|_{t=0}\,,
		\label{total_cross_section}
	\end{align}
	while the ratio of the real to the imaginary FSA is
	\begin{align}
		\rho=\frac{\text{Re}\,T(s,0)}{\text{Im}\,T(s,0)}\,.
		\label{rho_parameter}
	\end{align}

\subsection{The Bialas-Bzdak model with a unitarily extended amplitude}

	The original BB model~\cite{Bialas:2006qf} describes the proton as a bound state of a quark and a diquark, where both constituents have to be understood as ``dressed'' objects that effectively include all
	possible virtual gluons and $q\bar{q}$ pairs to  valence or dressed quarks. The quark and the diquark are characterized with their positions with respect to the proton's center of mass using their transverse position vectors $\vec{s}_{q}$ and $\vec{s}_{d}$ in the plane 
	perpendicular to the proton's incident momentum. Hence, the coordinate space~$H$ of the colliding protons is spanned by the vector $h=(\vec{s}_{q},\vec{s}_{d},\vec{s}^{\,\prime}_{q},\vec{s}^{\,\prime}_{d})$ where the
	primed coordinates indicate the coordinates of the second proton. 
	
	The inelastic proton-proton scattering probability $\tilde\sigma_{inel}(b)$ in Eq.~(\ref{original_BB_model}) is calculated as an average of ``elementary'' inelastic scattering probabilities $\sigma(h;{\vec b})$ over the 
	coordinate space $H$~\cite{Lipari:2013kta}
	\begin{equation}
	\tilde\sigma_{inel}(b)=\left<\sigma(h;{\vec b})\right>_{H}=\int\limits^{+\infty}_{-\infty}...\int\limits^{+\infty}_{-\infty}{{\rm d}h\,p(h)\cdot\sigma(h;{\vec b})}\,,
	\label{sigma_b_BB}
	\end{equation}
	where the weight function~$p(h)$ is a product of probability distributions
	\begin{equation}
		p(h)=D({\vec s}_q,{\vec s}_d)\cdot D({\vec s}^{\,\prime}_{q},{\vec s}^{\,\prime}_{d})\,.
		\label{probability_distribution}
	\end{equation}
	The $D({\vec s}_q,{\vec s}_d)$ function is a two-dimensional Gaussian, which describes the center of mass distribution of the quark and diquark with respect to the
	center of mass of the proton
    \begin{equation}
        D\left({\vec s}_q,{\vec s}_d\right)=\frac{1+\lambda^2}{R_{qd}^2\,\pi}e^{-(s_q^2+s_d^2)/R_{qd}^2}\delta^2({\vec s}_d+\lambda{\vec s}_q),\;\lambda=\frac{m_q}{m_d}\,.
	\label{quark_diquark_distribution}
    \end{equation}
	The parameter $R_{qd}$, the standard deviation of the quark and diquark distance, is fitted to the data. Note that the two-dimensional Dirac $\delta$ function preserves the proton's center of mass and reduces the dimension of the integral in Eq.~(\ref{sigma_b_BB}) from eight to four.

	Note that the original BB model is realized in two different ways: in one of the cases, the diquark structure is not
	resolved. This is referred to as the $p = (q,d)$ BB model. A more detailed variant is when the diquark is assumed to be a
	composition of two quarks, referred as the $p=(q,(q,q))$.
	Our earlier studies using the $\alpha$BB model
	indicated~\cite{CsorgO:2013kua}, that the $p = (q, d)$ case gives somewhat improved confidence levels
	as compared to the $p=(q,(q,q))$ case. So for the present manuscript we discuss results using the $p = (q,d)$ scenario
	only, however, it is trivial to extend the investigations to the $p=(q,(q,q))$ case and they result in fits which are
	not acceptable at $\sqrt{s}=7$~TeV. For the case of brevity we do not present the results of the analysis with the $p = (q, (q,q))$ variant of the
	ReBB model, only the fit quality is reported.

	It is assumed that the ``elementary'' inelastic scattering probability $\sigma(h;{\vec b})$ can be factorized in terms of binary collisions among the
	constituents with a Glauber expansion 
    \begin{align}
         \sigma(h;{\vec b})=1-\prod_{a}\prod_{b}\left[1-\sigma_{ab}({\vec b} + {\vec s}^{\,\prime}_{a} - {\vec s}_b )\right]\,,\quad a,b\in\{q,d\}\,,
    \label{Glauber_expansion}
    \end{align}
	where the indices~$a$ and~$b$ can be either quark~$q$ or diquark~$d$.

	The~$\sigma_{ab}\left({\vec s}\right)$ functions describe the probability of binary inelastic collision between quarks and diquarks and are assumed
	to be Gaussian
    \begin{equation}
        \sigma_{ab}\left({\vec s}\right) = A_{ab}e^{-s^2/S_{ab}^2},\;S_{ab}^2=R_a^2+R_b^2,\quad a,b \in \{q,d\}\,,
        \label{inelastic_cross_sections}
    \end{equation}
	where the $R_{q},R_{d}$ and $A_{ab}$ parameters are fitted to the data.

	The inelastic cross-sections of quark, diquark scatterings can be calculated by integrating the probability distributions~Eq.~(\ref{inelastic_cross_sections}) as
    \begin{equation}
    \label{totalinelastic}
        \sigma_{ab,\text{inel}}=\int\limits^{+\infty}_{-\infty}\int\limits^{+\infty}_{-\infty}{\sigma_{ab}\left({\vec s}\right)}\,\text{\rm d}^2s= \pi A_{ab}S_{ab}^2\,.
    \end{equation}
	In order to reduce the number of free parameters, it is assumed that the ratios of the inelastic cross-sections $\sigma_{ab,\text{inel}}$ satisfy
    \begin{equation}
        \sigma_{qq,\text{inel}}:\sigma_{qd,\text{inel}}:\sigma_{dd,\text{inel}}=1:2:4\,, 
        \label{ratiosforsigma}
    \end{equation}
	which means that in the BB model the diquark contains twice as many partons than the quark and also that these quarks and diquarks do not ``shadow'' each other during the scattering
	process. This assumption is not trivial. The $p=(q,(q,q))$ version of the BB model allows for different $\sigma_{qq,\text{inel}}:\sigma_{qd,\text{inel}}:\sigma_{dd,\text{inel}}$ ratios. However,
	as it was mentioned before, the $p=(q,(q,q))$ is less favored by the data as compared to the $p = (q,d)$ case presented below.

	Using the inelastic cross-sections~Eq.~(\ref{totalinelastic}) together with the assumption~Eq.~(\ref{ratiosforsigma}) the $A_{qd}$ and $A_{dd}$ parameters
	can be expressed with $A_{qq}$
        \begin{equation}
            A_{qd}=A_{qq}\frac{4R_q^2}{R_q^2+R_d^2}\,,\;A_{dd}=A_{qq}\frac{4R_q^2}{R_d^2}\,.
        \end{equation}
	In this way only five parameters have to be fitted to the data $R_{qd}$, $R_{q}$, $R_{d}$, $\lambda$, and $A_{qq}$. In practice we fix $A_{qq}=1$ assuming that head on quark-quark ($qq$)
	collisions are completely inelastic according to Eq.~(\ref{inelastic_cross_sections}).
 	
	The last step in the calculation is to perform the Gaussian integrals in the average~Eq.~(\ref{sigma_b_BB}) to obtain a formula for~$\tilde\sigma_{inel}(b)$. The Dirac $\delta$ function
	in~Eq.~(\ref{quark_diquark_distribution}) expresses the protons' diquark position vectors as a function of the quarks position
    \begin{equation}
        {\vec s}_d=-\lambda\,{\vec s}_q,\,\; {\vec s}^{\,\prime}_{d}=-\lambda\,{\vec s}^{\,\prime}_{q}\,.
	\label{Dirac_deltas}
    \end{equation}

	After expanding the products in the Glauber expansion~Eq.~(\ref{Glauber_expansion}) the following sum of contributions is obtained
 	\begin{align}
         \sigma(h;{\vec b})=&\sigma_{qq}+2\cdot\sigma_{qd}+\sigma_{dd}-(2\sigma_{qq} \sigma_{qd}+\sigma_{qd}^{2}+\sigma_{qq}\sigma_{dd}+2\sigma_{qd} \sigma_{dd})\notag\\
		&+(\sigma_{qq}\sigma_{qd}^{2}+2\sigma_{qq}\sigma_{qd}\sigma_{dd}+\sigma_{dd}\sigma_{qd}^{2})-\sigma_{qq}\sigma_{qd}^{2}\sigma_{dd}\,,
	\label{Glauber_expansion_2}
	\end{align}
	where the arguments of the $\sigma_{ab}(\vec{s})$ functions are suppressed to abbreviate the notation.

	The average over H in Eq.~(\ref{sigma_b_BB}) has to be calculated for each term in the above expansion Eq.~(\ref{Glauber_expansion_2}). Take the last, most general, term
	and calculate the average; the remaining terms are simple consequences of it. The result is
	\begin{equation}
		I=\left< -\sigma_{qq}\sigma_{qd}^{2}\sigma_{dd} \right>_{H}=\int\limits^{+\infty}_{-\infty}...\int\limits^{+\infty}_{-\infty}{{\rm d}h\,p(h)\cdot(-\sigma_{qq}\sigma_{qd}^{2}\sigma_{dd}\,)}\,,
		\label{sigma_b_BB_2}
	\end{equation}
	where the $p(h)$ weight function Eq.~(\ref{probability_distribution}) is a product of the quark-diquark distributions, given by Eq.~(\ref{quark_diquark_distribution}). Substitute into this result Eq.~(\ref{sigma_b_BB_2}) the definitions of the quark-diquark distributions Eq.~(\ref{quark_diquark_distribution})
    \begin{align}
		I=-\frac{4v^{2}}{\pi^2}\int\limits^{+\infty}_{-\infty}\int\limits^{+\infty}_{-\infty}{\rm d}^2s_q {\rm d}^2s_q'\,e^{-2v\left(s_q^2+s_q'^2\right)}\prod_{k}\prod_{l}\sigma_{kl}(\vec{b}-\vec{s}_{k}+\vec{s}_{l}^{\,\prime}),\quad k,l \in \{q,d\}\,,
            \label{master_formula_1}
    \end{align}
	where $v=(1+\lambda^{2})/(2\cdot R_{qd}^{2})$ and the integral over the coordinate space $H$ is explicitly written out; it is only four dimensional due to the two Dirac $\delta$ functions in $p(h)$. Using the definitions of the $\sigma_{ab}\left({\vec s}\right)$ functions Eq.~(\ref{inelastic_cross_sections}) and the 
	expression $A=A_{qq}A_{qd}A_{dq}A_{dd}$ the integral Eq.~(\ref{master_formula_1}) can be rewritten, to make all the Gaussian integrals explicit 
    \begin{align}
	I=-\frac{4v^{2}A}{\pi^2}\int\limits^{+\infty}_{-\infty}\int\limits^{+\infty}_{-\infty}{{\rm d}^2s_q {\rm d}^2s_q'\,e^{-2v\left(s_q^2+s_q'^2\right)}
            \prod_{k}\prod_{l}e^{-c_{kl}\left(\vec{b}-\vec{s}_{k}+\vec{s}_{l}^{\,\prime}\right)^2}}\,,
            \label{master_formula_2}
    \end{align}
	where the abbreviations $c_{kl}=S_{kl}^{-2}$ refer to the coefficients in Eq.~(\ref{inelastic_cross_sections}). Finally, the four Gaussian integrals have to be evaluated in our last expression Eq.~(\ref{master_formula_2}),
	which leads to
    \begin{align}
		I=-\frac{4v^{2}A}{B}e^{-b^2\frac{\Gamma}{B}}\,,
            \label{master_formula}
    \end{align}
	where
    \begin{align}
        B&=C_{qd,dq}\left(v+c_{qq} + \lambda^2 c_{dd}\right)+\left(1-\lambda\right)^2 D_{qd,dq}\,,\notag\\
        \Gamma&=C_{qd,dq}D_{qq,dd} + C_{qq,dd}D_{qd,dq}\,,
            \label{master_formula1}
    \end{align}
	and
    \begin{align}
	C_{kl,mn}&=4v + \left(1+\lambda\right)^2\left(c_{kl}+c_{mn}\right)\,,\notag\\
	D_{kl,mn}&=v \left(c_{kl}+c_{mn}\right)+\left(1+\lambda\right)^2c_{kl}c_{mn}\,.
	            \label{master_formula2}
    \end{align} 
	
	Each term in~Eq.~(\ref{sigma_b_BB}) can be obtained from the master formula Eq.~(\ref{master_formula}), by setting one or more coefficients to
	zero,~$c_{kl}=0$ and the corresponding amplitude to one,~$A_{kl}=1$.

	Up to now, according to~Eq.~(\ref{connection_omega_sigma}) and Eq.~(\ref{original_BB_model}), $t_{el}(s,b)$ is purely imaginary and $\Omega(s,b)$ is
	real. Now we have to specify the imaginary part of the opacity function, that determines the
	real part of the FSA. Here several model assumptions are possible, but from the
	analysis of the ISR data and the first studies of the 7 TeV TOTEM data at LHC we learned,
	that the real part of the FSA is perturbatively small at ISR energies, it becomes non-perturbative at
	LHC but the scattering is still dominated by the imaginary part of the scattering amplitude.

	We have studied several possible choices. One possibility is to introduce the imaginary
	part of the opacity function so that it is proportional to the probability of inelastic
	scatterings, which is known to be a decreasing function of the impact parameter b. A
	possible interpretation of this assumption may be that the inelastic collisions arising from
	non-collinear elastic collisions of quarks and diquarks follow the same spatial distributions
	as the inelastic collisions of the same constituents
	\begin{equation}
		\text{Im}\,\Omega(s,b)=-\alpha\cdot\tilde\sigma_{inel}(s,b)\,,
		\label{omega_b_with_sigma_b}
	\end{equation}
	where $\alpha$ is a real number.
	
	For the $\alpha = 0$ case, one recovers the $p=(q,d)$ version of the BB model of Ref.~\cite{Bialas:2006qf}, while in the $|\alpha|  \ll 1 $ perturbative limit
	the $\alpha$BB model of Ref.~\cite{CsorgO:2013kua} is obtained
	(but note the that the values of the parameter $\alpha$ in the two models need to be
	correspondingly re-scaled).

	The above proportionality between~$\text{Im}\,\Omega(s,b)$ and $\tilde\sigma_{inel}(b)$ in formula~(\ref{omega_b_with_sigma_b}) provided the best fits from among the relations that we have tried.
	For example, we have also investigated the assumption that the real and the imaginary parts of
	the opacity function are proportional to one another
	\begin{equation}
		\text{Im}\,\Omega(s,b)=-\alpha\cdot\text{Re}\,\Omega(s,b)\,.
		\label{omega_real_b_omega_imag_b}
	\end{equation}

 	However, as the results using Eq.~(\ref{omega_real_b_omega_imag_b}) were less favorable as the results obtained with Eq.~(\ref{omega_b_with_sigma_b}), we do the
	data analysis part,  described in the next section, using Eq.~(\ref{omega_b_with_sigma_b}).
	We mention this possibility to highlight that here some phenomenological assumptions are
	necessary as the ReBB model does allow for a broad range of
	possibilities for the choice of the imaginary part of the opacity function.

	In this way, the ReBB model is fully defined, and at a given colliding energy
	only six parameters determine the differential~(\ref{differential_cross_section}) and total cross-sections~(\ref{total_cross_section}) and
	also the $\rho$ parameter, defined with Eq.~(\ref{rho_parameter}).
	The parameters that have to be fitted to the data include the three scale parameters,
	$R_{q}$, $R_{d}$, $R_{qd}$,  that fix the geometry of the proton-proton collisions,
	as well as the three additional parameters $\alpha$, $\lambda$ and $A_{qq}$. 
	Two of the latter three can be fixed:  
	$\lambda = 0.5$ if the diquark is very weakly bound, 
	so that its mass is twice as large as that of the valence quark, while 
	$A_{qq} = 1$ suggests that head-on $qq$ collisions are inelastic with a probability of 1.
	Thus in the actual data analysis only four parameters
	are fitted to the data at each $\sqrt{s}$: the three scale parameters $R_{q}$, $R_{d}$ and  $R_{qd}$, 
	as well as the parameter $\alpha$. As we shall see, the parameter $\alpha$ will play a key role when describing the shape of the dip
	of the differential cross-sections of elastic $pp$ scatterings at LHC energies.

\section{Fit method and results}
\label{sec:fitdescription}

	The proton-proton elastic differential cross-section data measured by the LHC TOTEM experiment at $7$~TeV is a compilation of two subsequent measurements~\cite{Antchev:2011vs,Antchev:2013gaa}.
	The squared four-momentum transfer value $t_{sep}=-0.375$~GeV$^{2}$ separates the two data sets.\footnote{The squared four-momentum transfer value $t_{sep}$ separates the bin centers at the common boundary,
	the two bins actually overlap~\cite{Antchev:2011vs,Antchev:2013gaa}.} Note, that the two datasets were taken with two different settings of the machine optics of the LHC accelerator.

	The ReBB model, defined with Eq.~(\ref{differential_cross_section}), was fitted to the data at ISR energies and at LHC energy of $\sqrt{s}=7$~TeV. The relation between the imaginary part of $\Omega(s,b)$ and $\alpha$ is defined with Eq.~(\ref{omega_b_with_sigma_b}).
	In agreement with our previous investigations the $A_{qq}=1$ and $\lambda=\frac{1}{2}$ parameters can be kept constant, which reduces the number of free parameters to four $R_{qd},\,R_{q},\,R_{d}$ and $\alpha$.

\begin{figure}[H]
		\centering
		\includegraphics[width=0.9\linewidth]{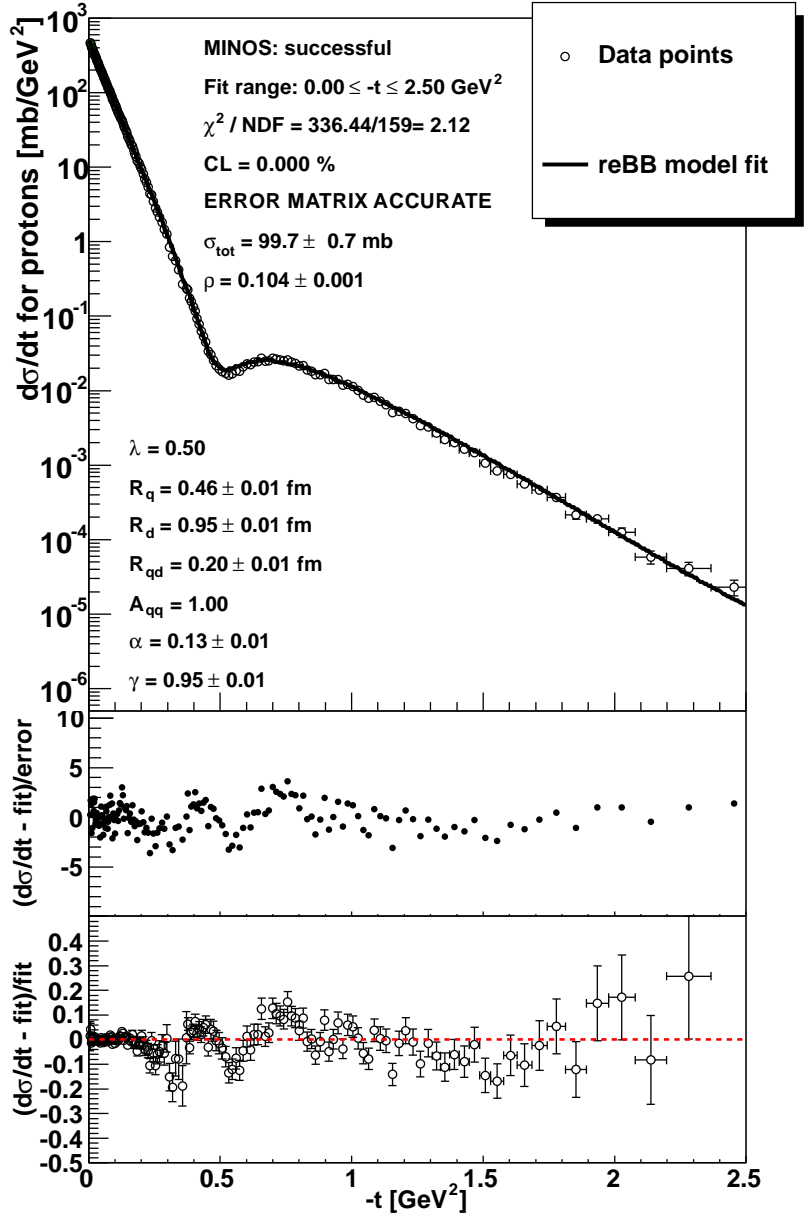}
		\caption{The fit of the ReBB model at $\sqrt{s}=7$~TeV in the $0 <|t| <2.5$~GeV$^{2}$ squared four momentum $|t|$ range. The real part of the amplitude $t_{el}$ is defined with
		expression Eq.~(\ref{omega_b_with_sigma_b}). According to Eq.~(\ref{differential_cross_section_gamma}) we use a relative normalization constant $\gamma$ between the
		two TOTEM datasets at $\sqrt{s}=7$~TeV. The fitted parameters are shown in the left bottom corner, parameters without errors were fixed in the minimization. The total cross-section~$\sigma_{tot}$
		and the parameter $\rho$ are derived quantities according to Eqs.~(\ref{total_cross_section}) and~(\ref{rho_parameter}), respectively.}
		\label{BBm_model_fit_results_fullb}
	\end{figure}

	First we have attempted to fit the ReBB model in the $0 < |t| < 2.5$ GeV$^{2}$ range, fitting simultaneously both the low-$|t|$ and the dip region.
	In the course of the minimization of the ReBB model at $\sqrt{s}=7$~TeV in this $t$-range, covering the two different TOTEM data sets, we found that the $\chi^{2}/NDF$ value decreases
	significantly, if a relative normalization constant $\gamma$ is introduced between the fit of the two data sets. Therefore, the calculated differential cross-section is fitted with
    	\begin{equation}
        	\frac{{\rm d}\sigma}{{\rm d}t}\rightarrow\gamma\cdot\frac{{\rm d}\sigma}{{\rm d}t}\,,
        	\label{differential_cross_section_gamma}
    	\end{equation}
    	if $|t|<|t_{sep}|$, where $\gamma$ is an additional parameter to be minimized. The fit at $\sqrt{s}=7$~TeV is shown in Fig.~\ref{BBm_model_fit_results_fullb}.

	Although the fit looks reasonable and reproducing the data qualitatively rather well,
	the fit quality it is not yet statistically acceptable, when the fit is extended to the
	whole $t$-region of the combined data set. Note that we determined the fit quality
	using statistical errors only, and as claimed in the original TOTEM publications~\cite{Antchev:2011vs,Antchev:2013gaa},
	the systematic errors in the two data set might be slightly different, that is
	rather difficult to handle correctly in the present analysis. So instead of determining
	the systematic errors of the model parameters from the systematic errors of the data
	we decided to analyze the two TOTEM data sets separately and check for the consistency
	of the results. As detailed below, this strategy lead to a reasonable fit qualities (CL = 2.6~\%, statistically acceptable
	fit in the cone region and CL = 0.04~\%, statistically marginal fit in the dip region) with
	a remarkable stability of fit parameters as detailed below.

	\begin{center}
	\begin{table}[H]\small
	\centering
		\begin{tabular}{|c|c|c|c|c|c|c|c|c|} \hline
		$\sqrt{s}$ [GeV] & 23.5                       	& 30.7                          & 52.8                          & 62.5          & \multicolumn{2}{|c|}{7000}\\   \hline\hline
		$|t|$ [GeV$^{2}$] & \multicolumn{4}{|c|}{$(0,2.5)$}							 			& $(0,|t_{sep}|$) 		& $(|t_{sep}|$,2.5)	\\   \hline
		$\chi^{2}/NDF$	&  124.7/102                   &   95.9/47            &     100.1/48      	&    76.6/47			& 109.9/83  		 			& 120.42/73	\\   \hline
		CL [\%] 	& 6.3                          & $3\times10^{-3}$              & $2\times10^{-3}$          	& 0.41    	& 2.6  		 		& $4\times10^{-2}$\\   \hline
		$R_{q}$ [$fm$] 	& 0.27$\pm$0.01   & 0.28$\pm$0.01     & 0.27$\pm$0.01     & 0.28$\pm$0.01     					& 0.45$\pm$0.01 		& 0.43$\pm$0.01	\\ \hline
		$R_{d}$ [$fm$] 	& 0.72$\pm$0.01   & 0.74$\pm$0.01     & 0.74$\pm$0.01     & 0.75$\pm$0.01     					& 0.94$\pm$0.01 		& 0.91$\pm$0.01	\\ \hline
		$R_{qd}$ [$fm$] & 0.30$\pm$0.01   & 0.29$\pm$0.01     & 0.33$\pm$0.01     & 0.32$\pm$0.01     					& 0.33$\pm$0.01 		& 0.37$\pm$0.02	\\ \hline
		$\alpha$ 	& 0.03$\pm$0.01   & 0.02$\pm$0.01     & 0.04$\pm$0.01     & 0.04$\pm$0.01     					& 0.12	  			& 0.12$\pm$0.01  \\ \hline
		\end{tabular}
		\caption{The values of the fitted ReBB model parameters.
		The proton-proton elastic $d\sigma/dt$ data measured by the TOTEM experiment at $7$~TeV is a composition of two subsequent measurements, which can be separated at $t_{sep}$.
		The overall fit involving the whole $0 <|t| <2.5$~GeV$^{2}$ range provides $\chi^{2}/NDF=336.4/159$ which is not statistically acceptable, while the fits below and above $|t_{sep}|$
		provide either a statistically acceptable (CL $>$ 0.1\%) or marginally good (CL = 0.04\%) fit quality.}
		\label{table:fit_parameters}
	\end{table}
	\end{center}

\begin{figure}[H]
		\centering
		\includegraphics[width=0.9\linewidth]{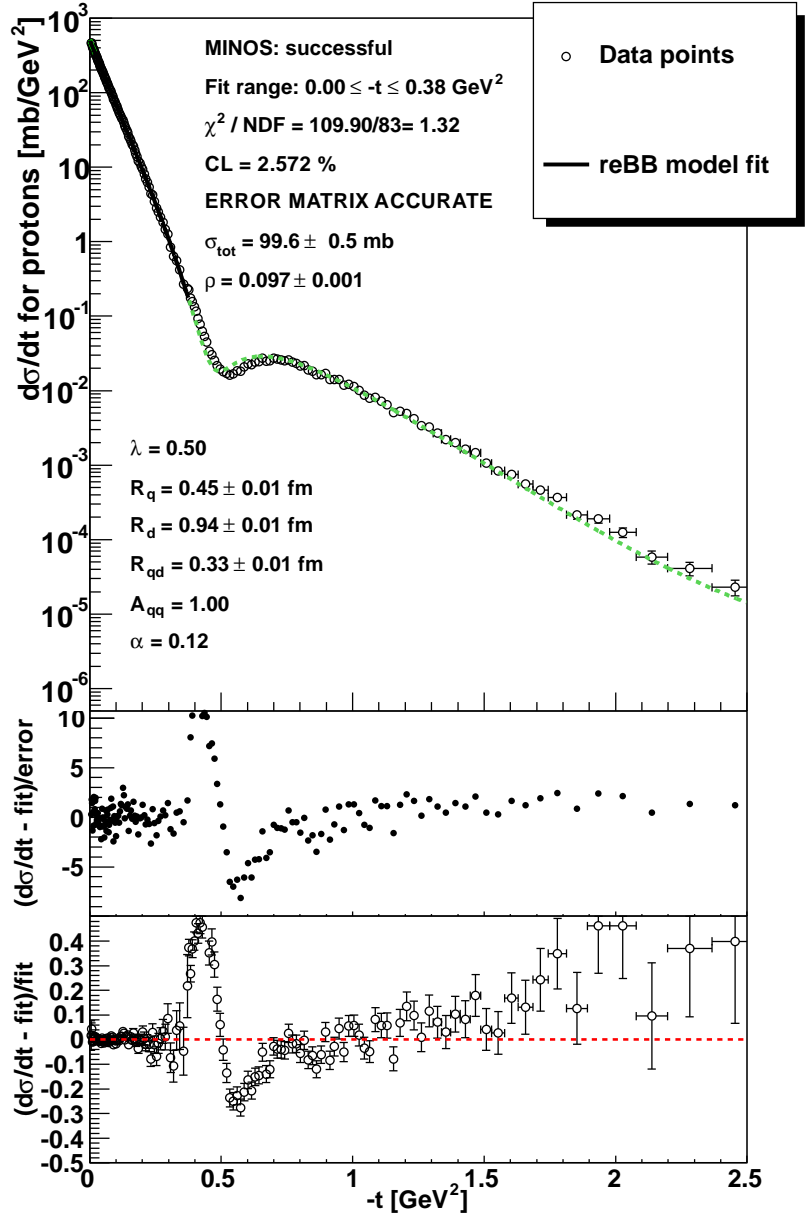}
		\caption{The same as Fig.~\ref{BBm_model_fit_results_fullb}, but the fit is evaluated in the $0<|t|<|t_{sep}|$ range. The
		fitted curve is shown with solid line, its extrapolation above $|t_{sep}|$ is indicated with a dashed line. Note that the extrapolated
		curve remains close to the data points,
		following the measured differential cross-sections well even far away from the region where the model was fitted to the data.}
		\label{BBm_model_fit_results_full}
	\end{figure}

\begin{figure}[H]
		\centering
		\includegraphics[width=0.9\linewidth]{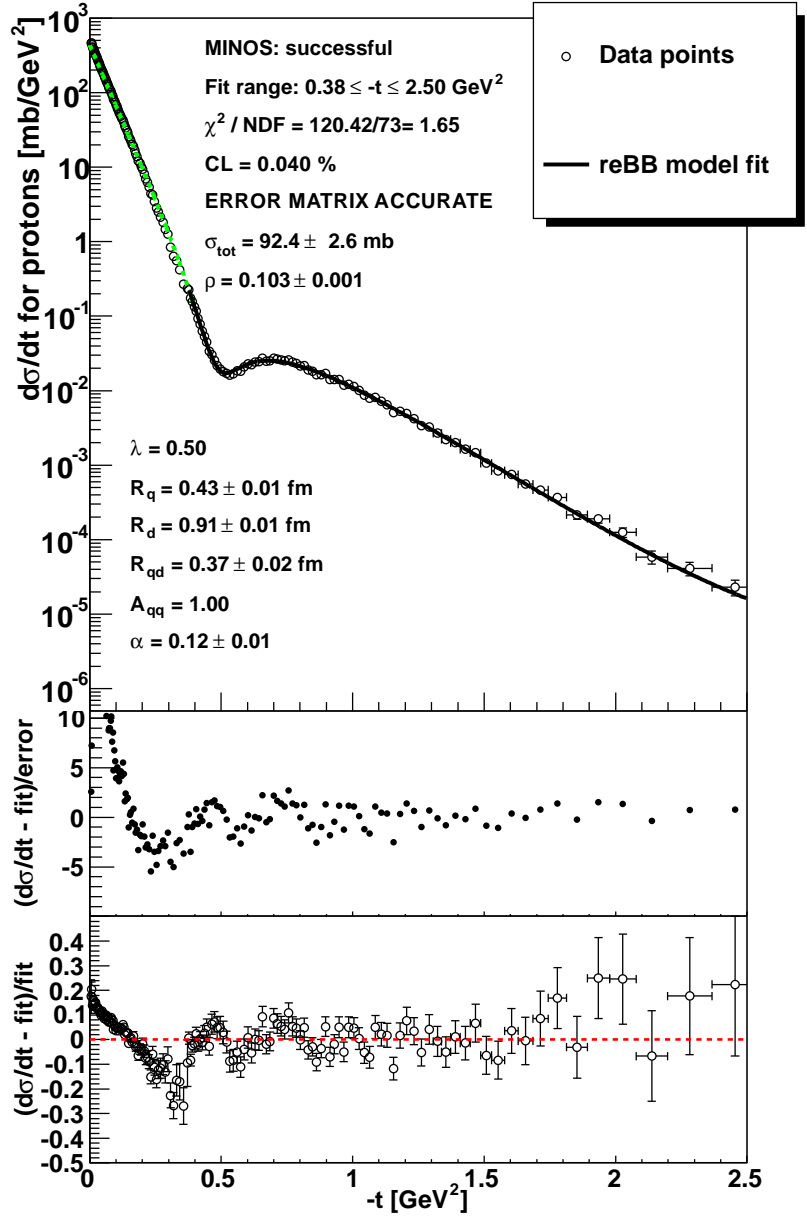}
		\caption{The same as Fig.~\ref{BBm_model_fit_results_fullb}, but the fit is performed in the $|t_{sep}|<|t|<2.5$~GeV$^{2}$ range. The
		fitted curve is shown with solid line, its extrapolation is indicated with a dashed line. Note that when the curve is extrapolated to the low-$|t|$ region,  the extrapolated curve again follows
		the measured differential cross-section remarkably well even far away the fit region: the ReBB model fit is
		remarkably stable over the whole $|t|$-range.} 
		\label{BBm_model_fit_results_fulla}
	\end{figure}

	If a {\it separated} fit to $\sqrt{s}=7$ TeV elastic differential cross-section $d\sigma/dt$ data is evaluated, below and above the separation $|t_{sep}|$, a quality result can be obtained, which is
	shown in Figs.~\ref{BBm_model_fit_results_full} and~\ref{BBm_model_fit_results_fulla}
	and reported in Table~\ref{table:fit_parameters}, together with our results at ISR energies~\cite{Antchev:2011vs,Antchev:2013gaa,Nagy:1978iw,Amaldi:1979kd}. Note that the  
	normalization factor $\gamma$, introduced in Eq.~(\ref{differential_cross_section_gamma}), is not applied at $\sqrt{s}=7$~TeV, as the two data set were fitted separately. 

	The resulting parameters coming from the two separate fits at 7 TeV $d\sigma/dt$, over and below the $|t_{sep}|$ value, are {\it consistent} with each other within $2\sigma$ error.
	Note that at $\sqrt{s}=7$~TeV the dip is not part of the fit range $(0,|t_{sep}|$), thus the minimization procedure cannot determine the value of parameter $\alpha$. In this case we have
	fixed $\alpha$ to the value of the fit from the other $|t|$ range above $|t_{sep}|$. The MINUIT status of the fit is successful in both cases.

	Due to the stability of the fit parameters the extrapolation of the fit curves to the not fitted $|t|$ range remains close to the data points. The stability and consistency
	of the model description is visible in Fig.~\ref{BBm_model_fit_results_full} and~\ref{BBm_model_fit_results_fulla}.

	The calculated total cross-section of the low-$|t|$ fit $\sigma_{tot}=99.6\pm0.5$~mb, where the uncertainty is the propagated uncertainty of the fit parameters, agrees well with the value
	$\sigma_{tot}=98.0\pm2.5$~mb measured by the TOTEM experiment at $\sqrt{s}=7$~TeV~\cite{Antchev:2013iaa}.

	The parameter $\rho$ can be better estimated from the fit over $|t_{sep}|$ which includes the dip. As the measured value of the $\rho$ parameter $\rho= 0.145\pm0.091$ has large uncertainty
	the $\rho=0.103\pm0.001$ calculated from the ReBB model is consistent with the measurement, see Fig.~\ref{BBm_model_fit_results_full}.

	Also note that if $\text{Im}\,\Omega(s,b)$ is defined to be proportional to $\text{Re}\,\Omega(s,b)$, according to Eq.~(\ref{omega_real_b_omega_imag_b}), the MINUIT fit result of $\chi^{2}/NDF=405.6/159=2.55$ is obtained at $\sqrt{s}=7$~TeV,
	which is disfavored as compared to fits with Eq.~(\ref{omega_b_with_sigma_b}).

	In our introduction we shortly mentioned the $p=(q,(q,q))$ version of the ReBB model, when the diquark is assumed to be a composition of two quarks~\cite{CsorgO:2013kua}.
	This scenario provides a fit results with $\chi^{2}/NDF=15509/159\approx97.5$, which means that the $p=(q,(q,q))$ ReBB version can be clearly rejected. The failure of this version is basically due the wrong
	shape of the differential cross-section: the second diffractive minimum appears too close to the first one.

	\begin{figure}[h]
		\centering
		\includegraphics[width=0.48\linewidth]{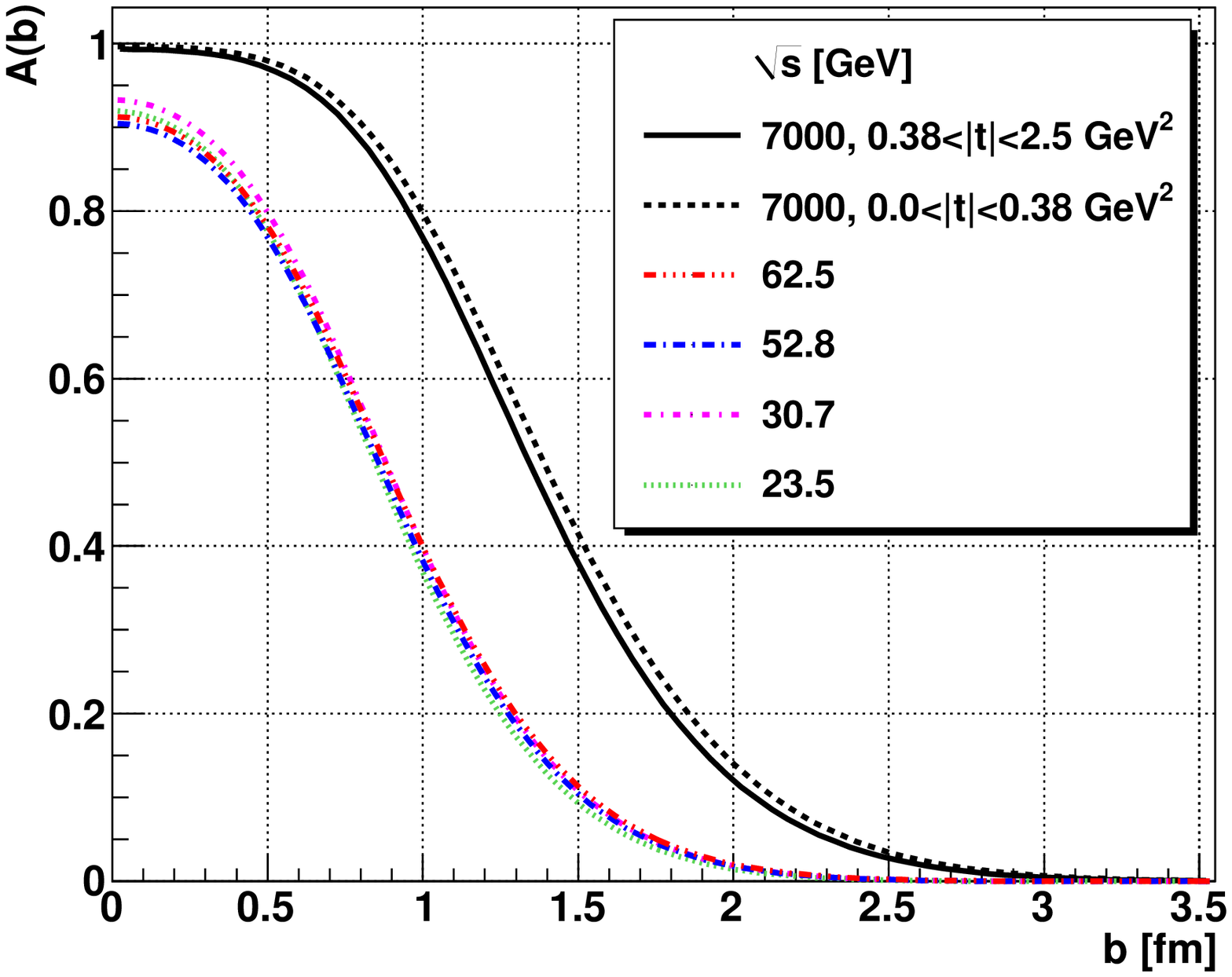}
		\includegraphics[width=0.48\linewidth]{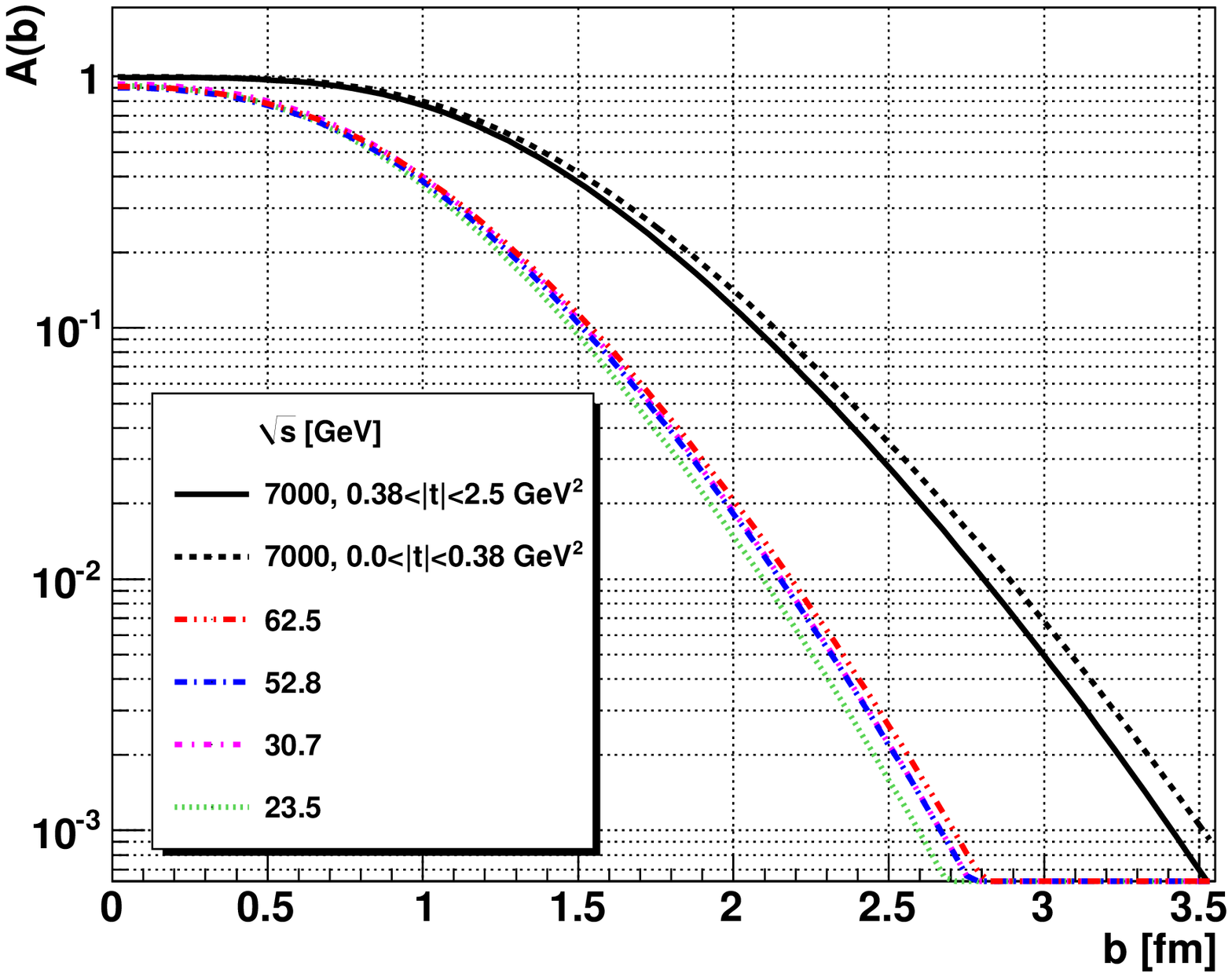}
		\caption{The shadow profile functions $A(b)$ indicate a saturation effect at LHC, while at ISR energies a Gaussian shape can be observed. Note that the dashed black curve is based on the statistically
		acceptable fit result in the $0<|t|<0.38$~GeV$^{2}$ range. The distributions' edge shows approximately the same width at each energy, corresponding to a constant ``skin-width'' of the proton.}
		\label{BBm_model_fit_results_shadow}
	\end{figure}

\section{Discussion}
\label{sec:discussion}
\subsection{Shadow profile functions and saturation}
\label{sec:shadow_profile}

	The fits, from which the model parameters were determined, also permit us to evaluate the shadow profile function 
	\begin{equation}
	 	A(s,b)=1-\left|\exp\left[-\Omega(s,b)\right]\right|^{2}\,.
	\end{equation}
	The obtained curves to $A(b)$ are shown in Fig.~\ref{BBm_model_fit_results_shadow}. The shadow profile functions at ISR energies exhibit a Gaussian like shape, which
	smoothly change with the center of mass energy $\sqrt{s}$. At LHC something new appears: the innermost part of the distribution shows a saturation, which means that
	around $b=0$ the function becomes almost flat and stay close to $A(b)\approx1$. Consequently, the shape of the shadow profile function $A(b)$ becomes non-Gaussian and somewhat ``distorted'' with respect to the shapes found
	at ISR. 
	
	At the same time the width of the edge of the shadow profile function A(b), which can be visualized as the proton's ``skin-width'', remains approximately independent of the center of mass energy $\sqrt{s}$.

\subsection{Non-exponential behavior of $d\sigma_{el}/dt$}
\label{sec:non_exponential}

	To compare the obtained ReBB fit with a {\it purely} exponential distribution the following exponential parametrization is used
	\begin{equation}
		\frac{\rm{d}\sigma_{el}}{\rm{d}t}=\left.\frac{\rm{d}\sigma_{el}}{\rm{d}t}\right|_{t=0}\cdot e^{-B\cdot|t|}\,,
		\label{exp_distribution_from_Jan}
	\end{equation}
	where $\left.\rm{d}\sigma_{el}/\rm{d}t\right|_{t=0}=506.4$~mb/GeV$^{2}$ and slope parameter $B=19.89$~GeV$^{-2}$ is applied, according to the TOTEM paper Ref.~\cite{Antchev:2013gaa}.

\begin{figure}[h]
		\centering
		\includegraphics[width=0.8\linewidth]{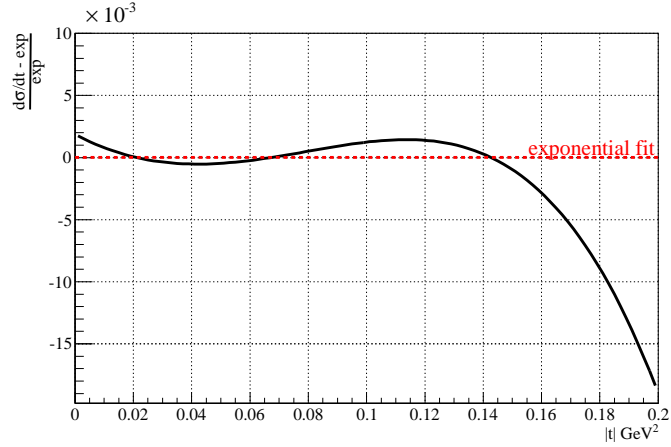}
		\caption{The ReBB model, fitted in the $0.0\le|t|\le0.36$~GeV$^{2}$ range, with respect to the exponential fit of Eq.~(\ref{exp_distribution_from_Jan}). In
		the plot only the $0.0\le|t|\le0.2$~GeV$^{2}$ range is shown. The curve indicates a significant deviation from the simple exponential at low $|t|$ values.}
		\label{BBm_model_fit_results_non_exponential}
	\end{figure}

	The result, shown in~Fig.~\ref{BBm_model_fit_results_non_exponential}, indicates a clear non-exponential behavior of the elastic differential cross-section in the $0.0\le|t|\le0.2$~GeV$^{2}$ range at $\sqrt{s}=$7 TeV.
	Note that a similar non-exponential behavior was recently discussed by the TOTEM experiment~\cite{Simone:WPCF2014} and also by the theoretical work
	of Ref.~\cite{Jenkovszky:2014yea}.

\section{Extrapolation to future LHC energies and beyond}
\label{sec:excitation}
\begin{figure}[H]
		\centering
		\includegraphics[width=0.495\linewidth]{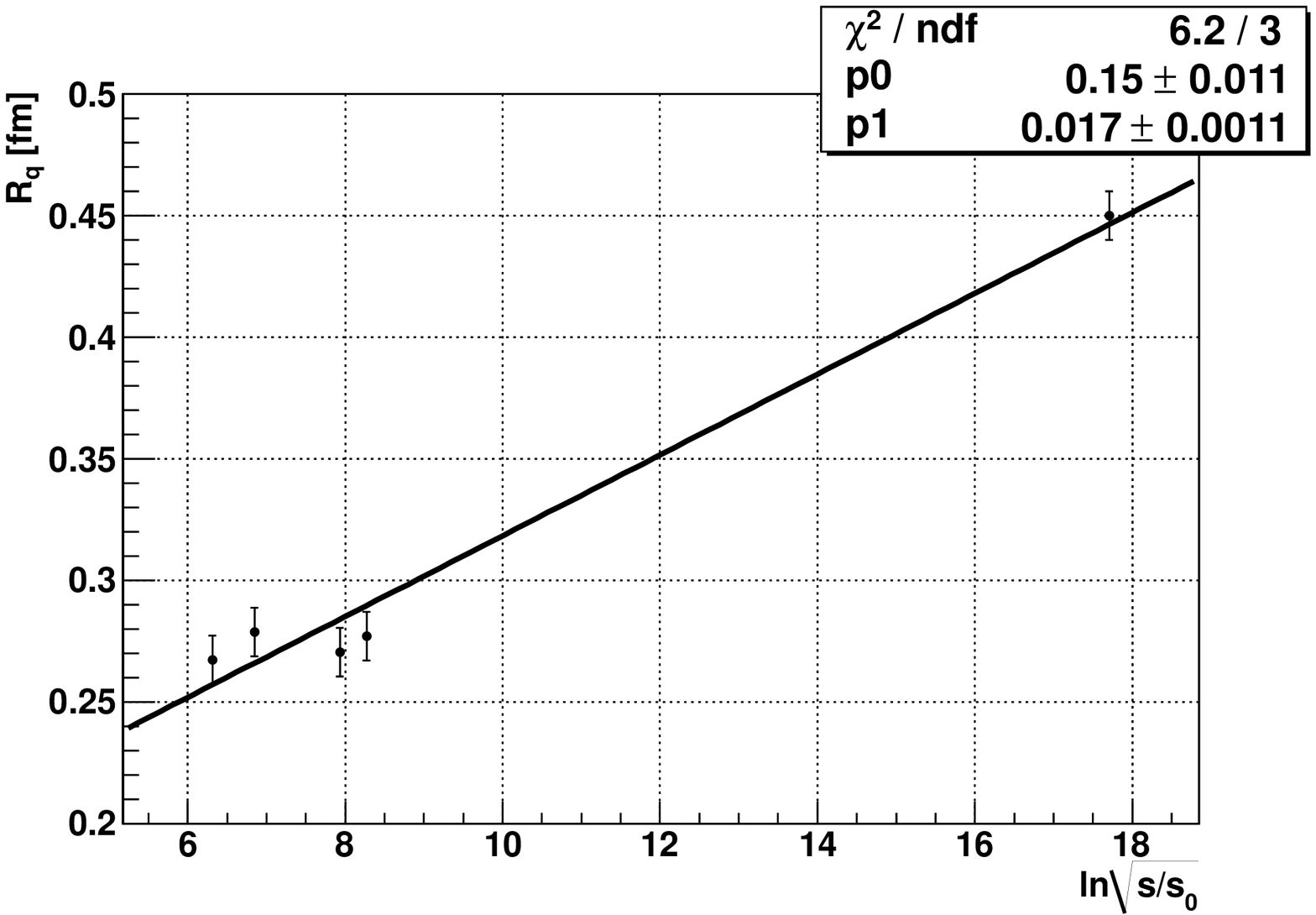}
		\includegraphics[width=0.495\linewidth]{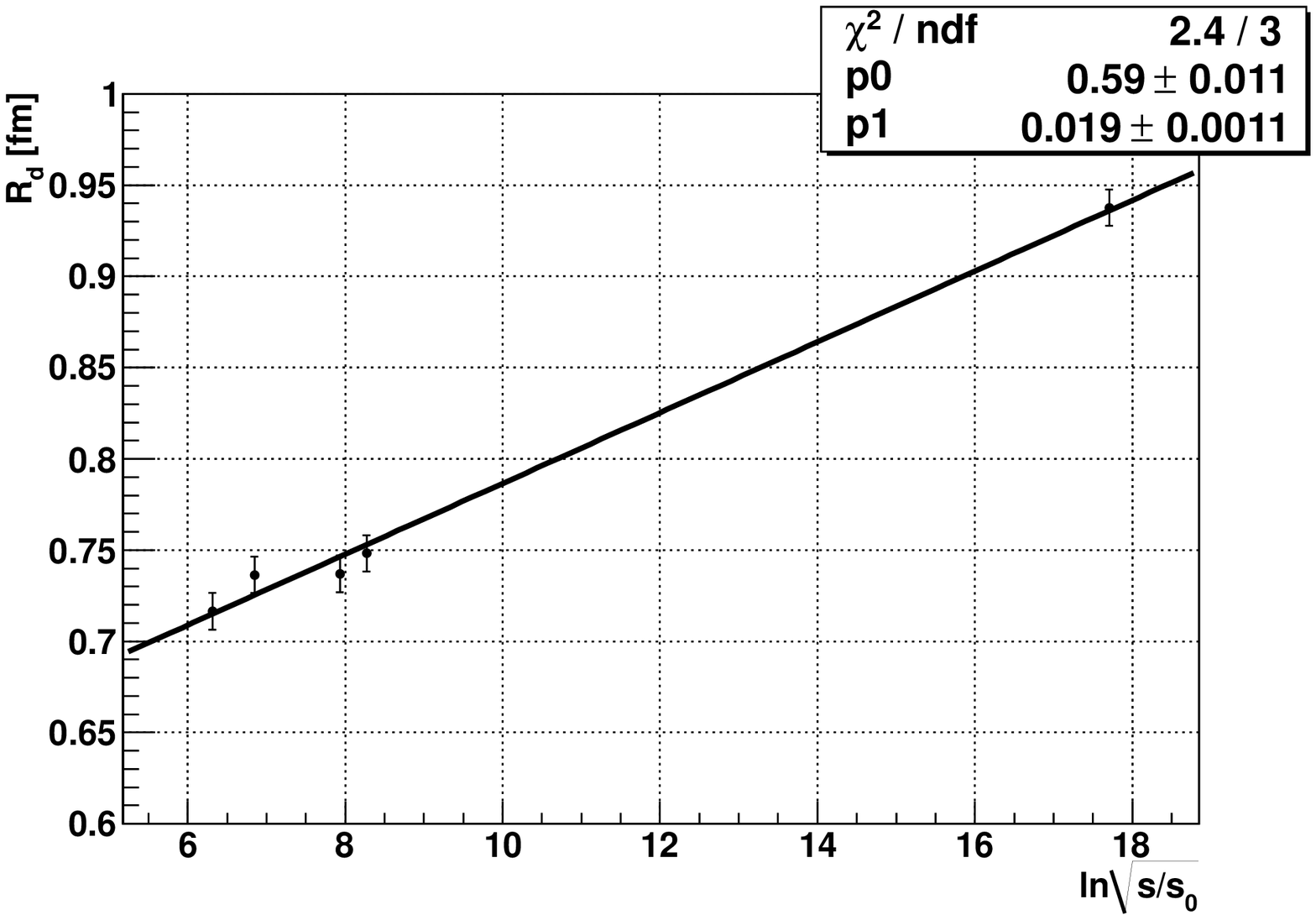}\\
		\includegraphics[width=0.495\linewidth]{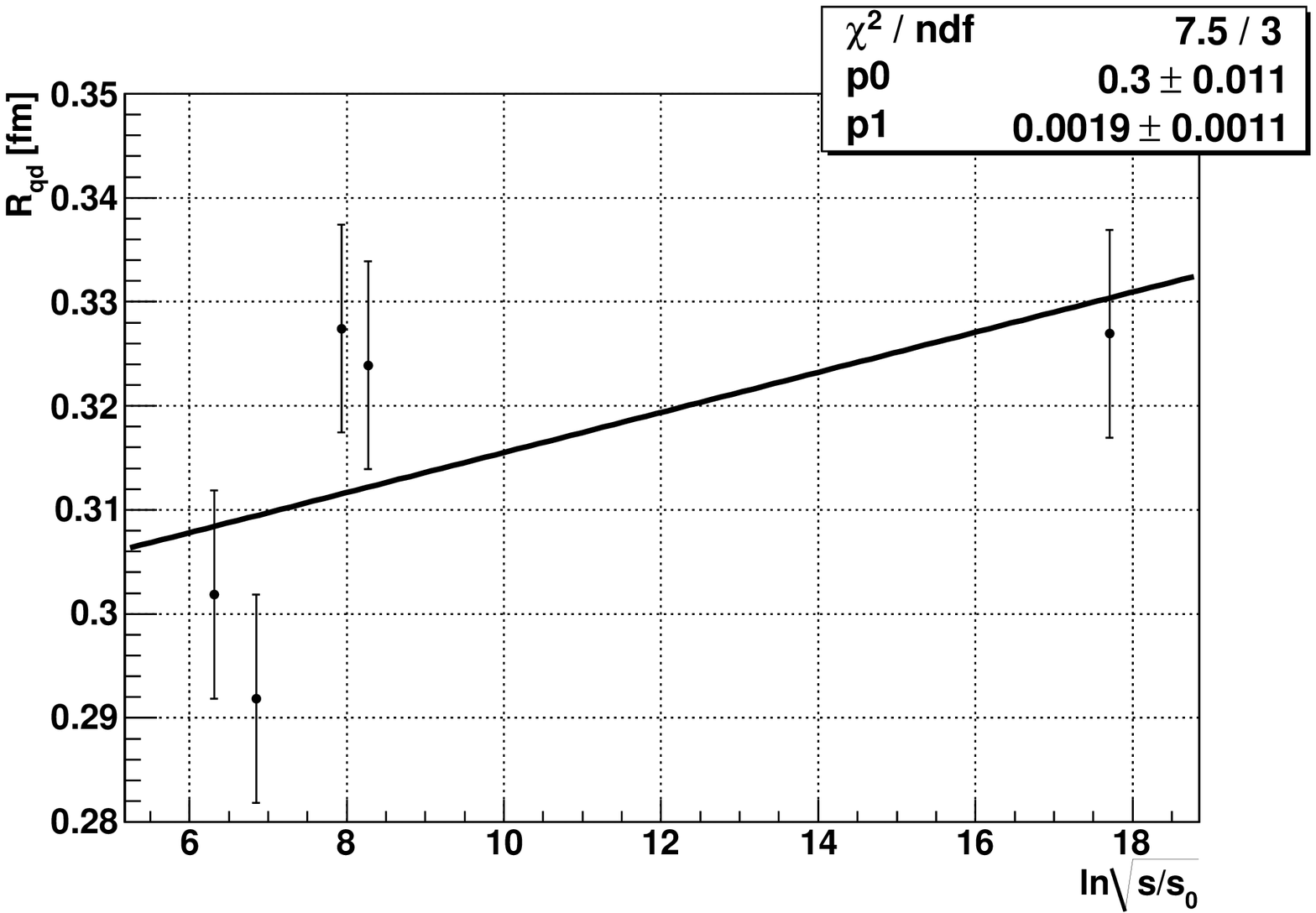}
		\includegraphics[width=0.495\linewidth]{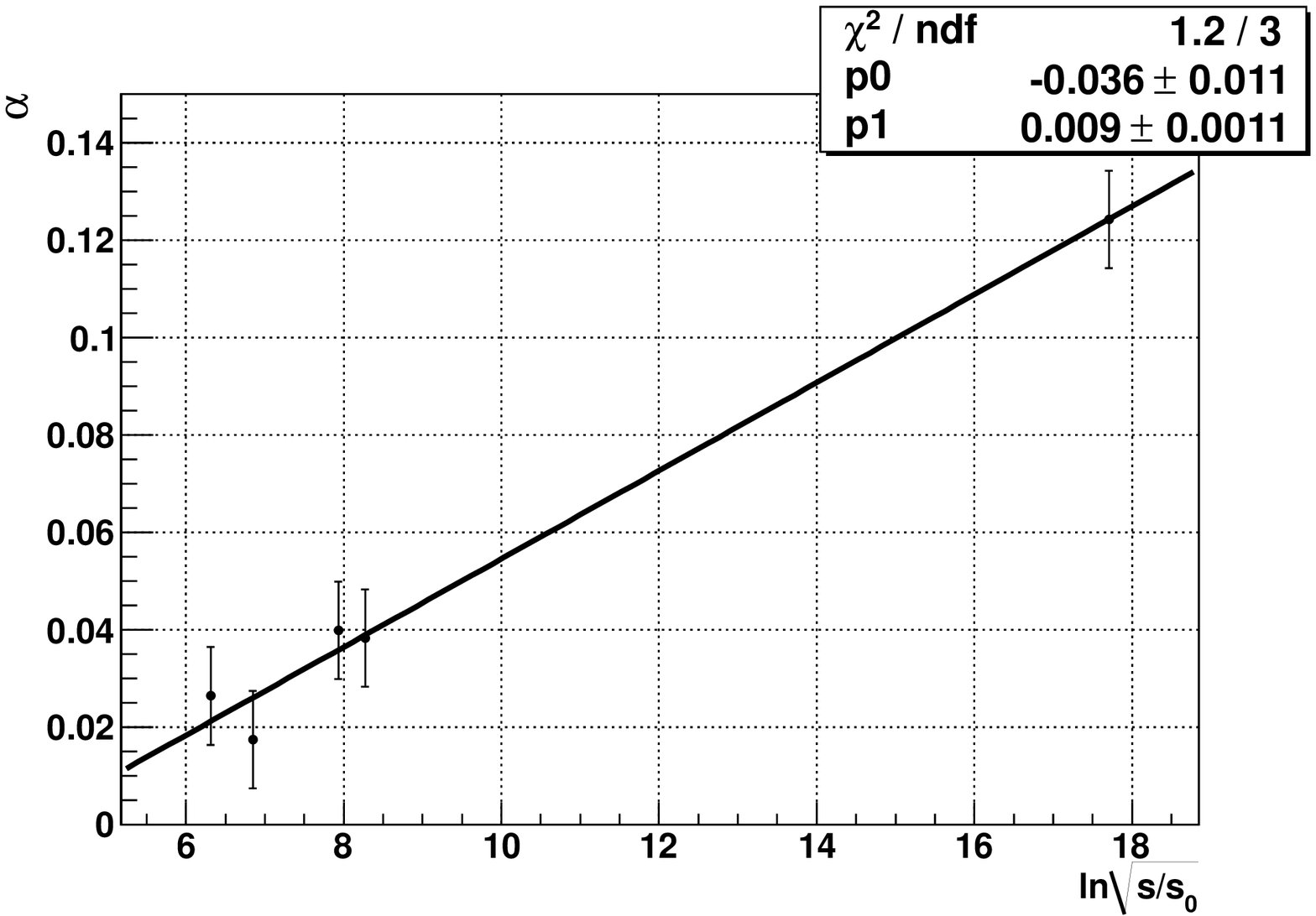}
		\caption{The results, collected in Table \ref{table:fit_parameters}, are fitted with Eq.~(\ref{parametrization_of_extrapolation}) for each 
		parameter $R_{q}$, $R_{d}$, $R_{qd}$ and $\alpha$. The plots about the resulting fits are collected here, the parameters are collected 
		in Table~\ref{extrapolation_parameters}. The statistically acceptable quality of these fits allow the ReBB
		model to be extrapolated to center of mass energies which have not been measured yet at LHC.}
		\label{BBm_model_extrapolation_fits}
	\end{figure}

	The ReBB model can be extrapolated to energies which have not been measured yet at LHC. The fit results of Table \ref{table:fit_parameters} and the parametrization  
	\begin{equation}
	 P(s)=p_{0} + p_{1}\cdot\ln{(s/s_{0})}
		\label{parametrization_of_extrapolation}
	\end{equation}
	is applied for each parameter $P\in{\{R_{q},R_{d},R_{qd},\alpha\}}$, where $s_{0}=1$ GeV$^{2}$. The parametrization Eq.~(\ref{parametrization_of_extrapolation}) implies that the four free parameters of the original
	ReBB model are replaced with eight parameters $p_{i}$. The fit of the ReBB parameters are shown in Fig.~\ref{BBm_model_extrapolation_fits} and the
	fit parameters are collected in Table~\ref{extrapolation_parameters}.
	The logarithmic dependence of the geometric parameters on the center of mass energy $\sqrt{s}$ in the parametrization Eq.~(\ref{parametrization_of_extrapolation}) is motivated by
	the so-called ``geometric picture`` based on a series of studies~\cite{Cheng:1969eh,Cheng:1969bf,Cheng:1969ac,Chengbook,Bourrely:1978da,Bourrely:2014efa}.

	Table~\ref{extrapolation_parameters} shows that the rate of increase with $\sqrt{s}$, parameter $p_{1}$, is an order of magnitude larger
	for $R_{q}$ and $R_{d}$ than for $R_{qd}$. The saturation effect, described in Section~\ref{sec:shadow_profile}, is consistent with this 
	observation as the increasing components of the proton, the quark and the diquark, are confined into a volume which is increasing more slowly.
	\begin{table}[H]
	\centering
		\begin{tabular}{|c|c|c|c|c|} \hline
		Parameter 	& $R_{q}$ [$fm$]                & $R_{d}$ [$fm$]       	& $R_{qd}$ [$fm$]	& $\alpha$		\\   \hline\hline
		$\chi^{2}/NDF$ 	& $6.2/3$			& $2.4/3$		& $7.5/3$		& $1.2/3$ 		\\   \hline
		CL [\%]		& 10.2				& 49.4			& 5.8			& 75.3	 		\\   \hline
		$p_{0}$ 	& $0.15\pm0.01$			& $0.59\pm0.01$		& $0.3\pm0.01$		& $-0.036\pm0.01$ 	\\   \hline
		$p_{1}$ 	& $0.017\pm0.001$		& $0.019\pm0.001$	& $0.0019\pm0.001$	& $\phantom{-}0.009\pm0.001$ 	\\   \hline
		\end{tabular}
		\caption{Table \ref{table:fit_parameters} allows the extrapolation of the model parameters over the center of mass energy $\sqrt{s}$. The
		parametrization Eq.~(\ref{parametrization_of_extrapolation}) is applied to extrapolate the ReBB model and the fits are
		shown in Fig.~\ref{BBm_model_extrapolation_fits}. The fit quality information is provided in the first and second row
		of the table. Note that the fit quality is acceptable for each parameter.}
		\label{extrapolation_parameters}
	\end{table}
	\vspace{-5mm}
	\begin{wrapfigure}{r}{0.5\textwidth}
		\vspace{-30pt}
  		\begin{center}
    			\includegraphics[trim = 0mm 8mm 0mm 20mm, clip, width=0.5\textwidth]{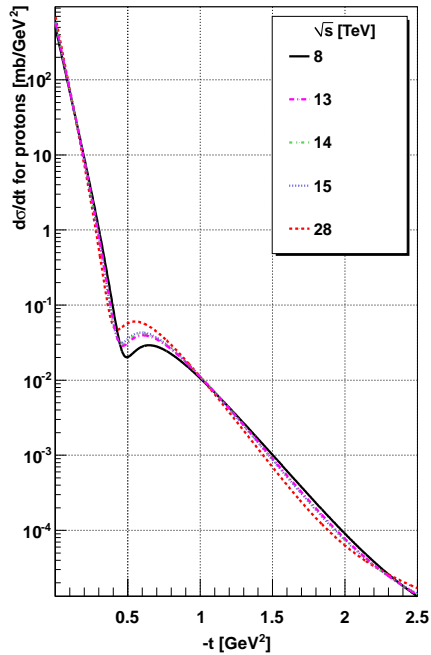}
  		\end{center}
  		\caption{The $pp$ elastic differential cross-section is extrapolated to future LHC energies and beyond.}
		\label{BBm_model_extrapolation}
		\vspace{-30pt}
	\end{wrapfigure}
	
\begin{figure}[H]
		\centering
		\includegraphics[width=0.495\linewidth]{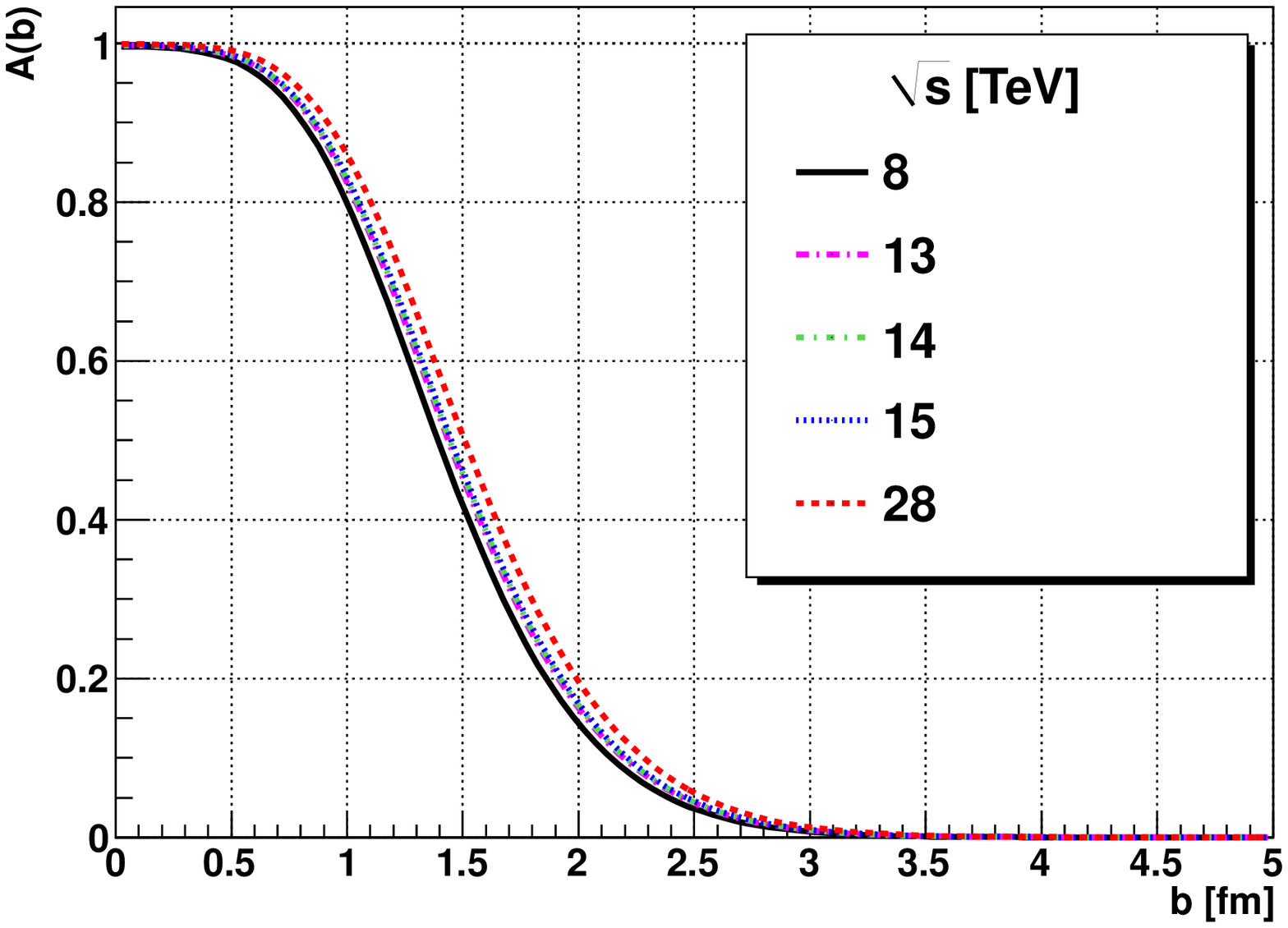}
		\includegraphics[width=0.495\linewidth]{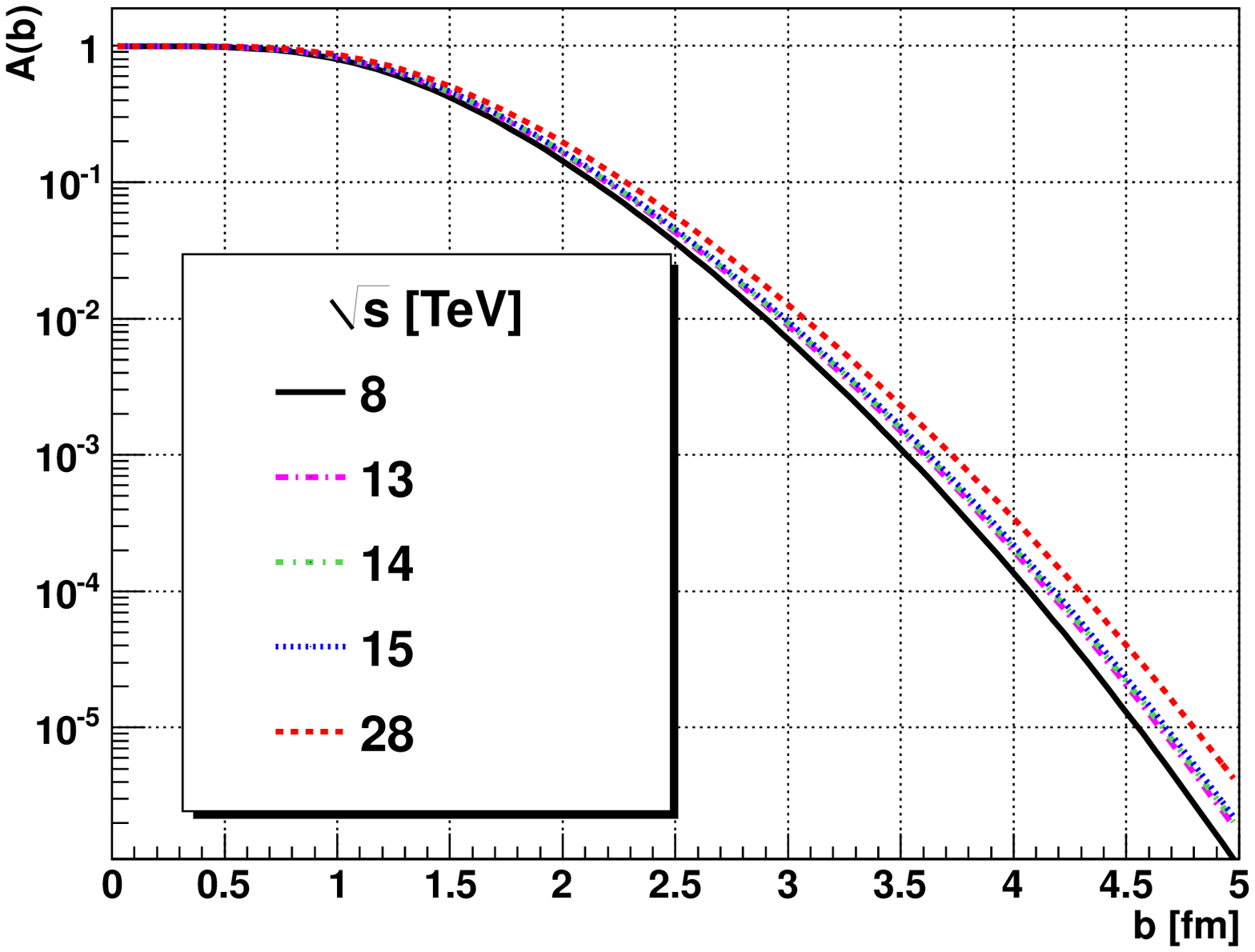}
		\caption{The shadow profile function at the extrapolated energies $\sqrt{s}$. The results show the increase of the proton interaction radius with increasing $\sqrt{s}$ energies. Also
		note that the ``edge'' of the distributions remains of approximately constant width and shape.}
		\label{BBm_model_extrapolation_shadows}
	\end{figure}

\begin{figure}[h]
		\centering
		\includegraphics[width=0.49\linewidth]{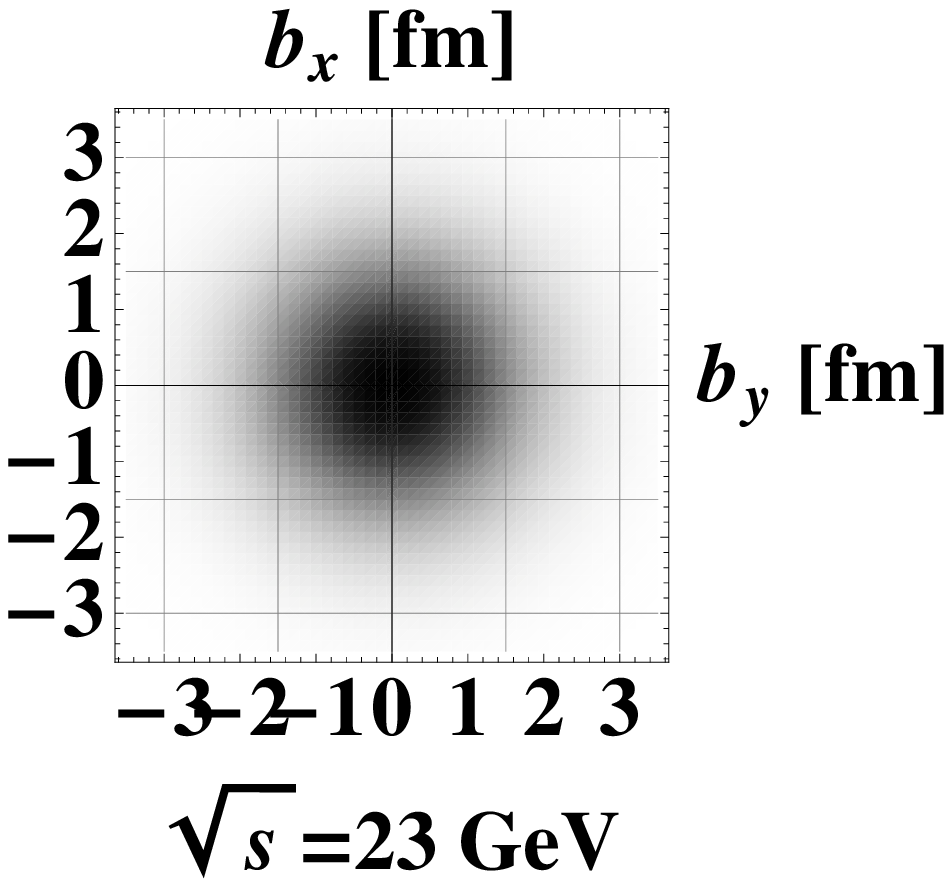}
		\includegraphics[width=0.49\linewidth]{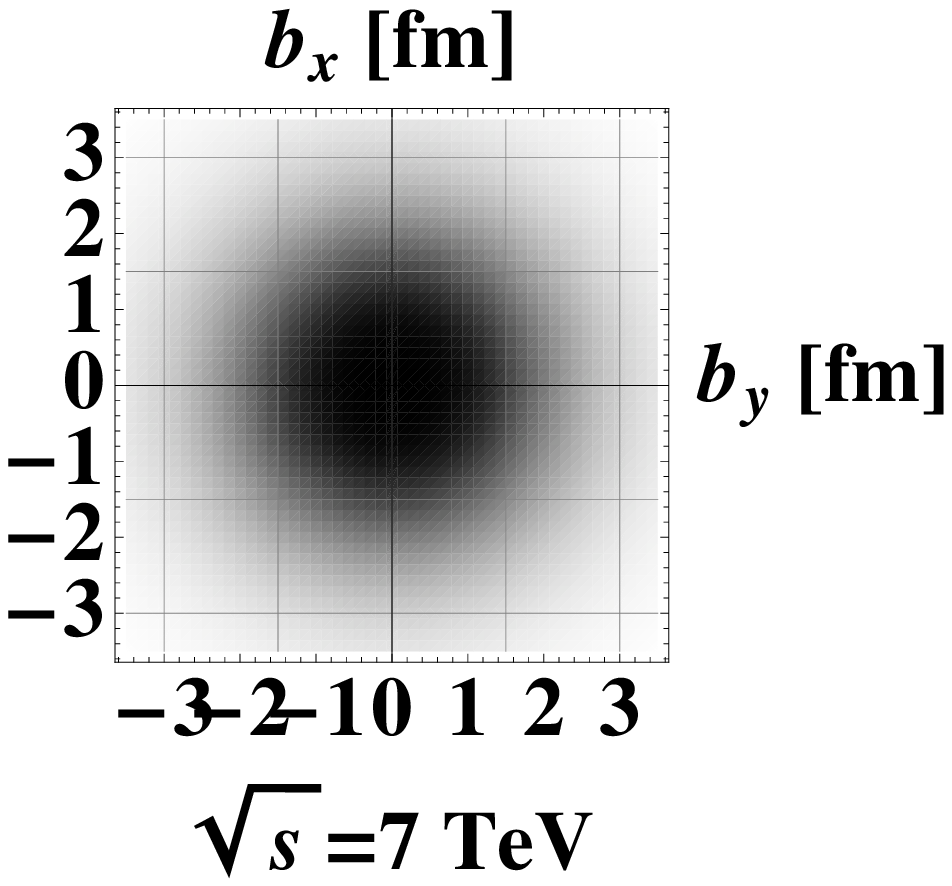}
		\includegraphics[width=0.49\linewidth]{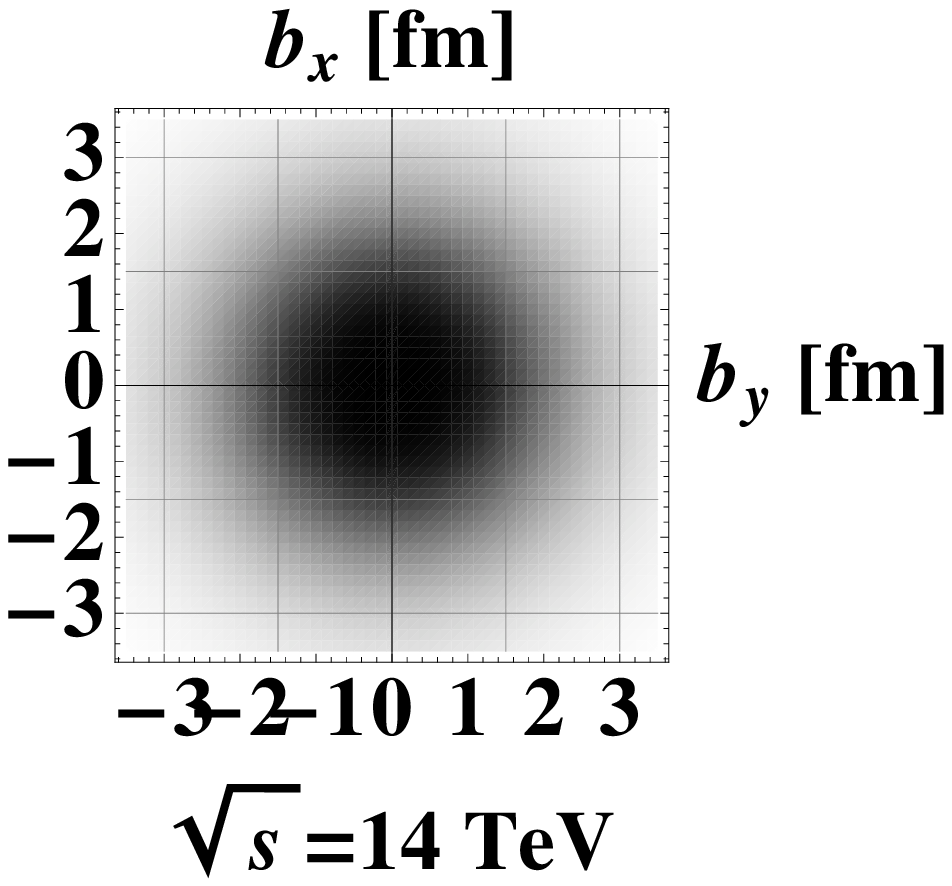}
		\includegraphics[width=0.49\linewidth]{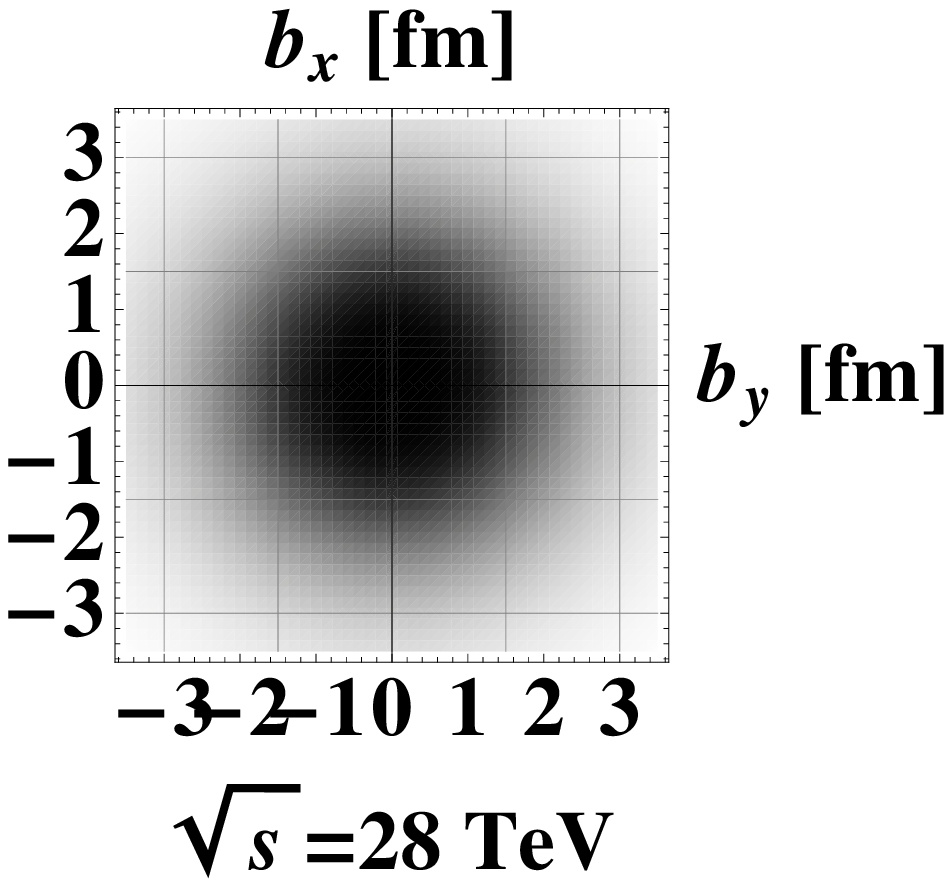}
		\caption{Visualization of the shadow profile functions $A(b)$ in the transverse plane of the impact parameter vector~$(b_{x},b_{y})$. The figures show the increase of the proton 
		effective interaction radius in the impact parameter space with increasing center of mass energy~$\sqrt{s}$. It can be also observed that the black innermost core of the distributions
		is increasing, while the thickness of the proton's ``skin'', the gray transition part of the distributions, remains approximately independent of the center of mass energy $\sqrt{s}$.}
		\label{BBm_model_fit_results_shadow_visualization}
	\end{figure}

	Using the extrapolation formula Eq.~(\ref{parametrization_of_extrapolation}) and the value of the parameters from Table~\ref{extrapolation_parameters} it is straightforward to calculate
	the values of the parameters at expected future LHC energies of $\sqrt{s}=$8, 13, 14, 15 TeV and also at 28 TeV, which is beyond the LHC capabilities. Using the extrapolated values of the parameters we plot our
	predicted $pp$ elastic differential cross-section curves at each mentioned energy in Fig.~\ref{BBm_model_extrapolation}.
	The shadow profile functions $A(b)$ can be also extrapolated, see Fig.~\ref{BBm_model_extrapolation_shadows}.
	The shadow profile functions even allow us to visualize the increasing effective interaction radius of the proton in the impact parameter space in Fig.~\ref{BBm_model_fit_results_shadow_visualization}. 

	It is also important to see how the most important features change with center of mass energy~$\sqrt{s}$: the extrapolated values of the
	total cross-section $\sigma_{tot}$, the position of the first diffractive minimum $|t_{dip}|$ and the parameter $\rho$ is
	given in Table \ref{extrapolated_values}.

	According to the results, the predicted value of $|t_{dip}|$ and $\sigma_{tot}$ moves more than 10\% when $\sqrt{s}$ increases from
	8~TeV to 28~TeV, while the value of $C_{exp}=|t_{dip}|\cdot\sigma_{tot}\approx49.8$~mb~GeV$^{2}$ changes only about 2~\%, which is an approximately constant value, within the errors of the extrapolation.

	\begin{table}[H]
	\centering
		\begin{tabular}{|c||c|c|c|c|} \hline
		$\sqrt{s}$ [TeV] & $\sigma_{tot}$ [mb] & $|t_{dip}|$[GeV$^{2}$] & $\rho$ & $|t_{dip}|\cdot\sigma_{tot}$ [mb~GeV$^{2}$] \\   \hline\hline
			8	 & 100.1	& 0.494 & 0.103	& 49.45 \\   \hline
			13	 & 107.1	& 0.465 & 0.108	& 49.8 \\   \hline
			14	 & 108.1 	& 0.461 & 0.108	& 49.83 \\   \hline
			15	 & 109.1	& 0.457 & 0.109	& 49.86 \\   \hline
			28	 & 118.5	& 0.426 & 0.114	& 50.48 \\   \hline
		\end{tabular}
		\caption{The extrapolated values of the total cross-section $\sigma_{tot}$ at future LHC energies and beyond. The position of the first diffractive minimum $|t_{dip}|$, the
		parameter $\rho$ and the $|t_{dip}|\cdot\sigma_{tot}$ value is also provided at each energy. Note that the predicted value of $|t_{dip}|$ and $\sigma_{tot}$ moves more than 10\% when
		$\sqrt{s}$ increases from 8~TeV to 28~TeV, while the value of $|t_{dip}|\cdot\sigma_{tot}$ changes only about 2\%.}
		\label{extrapolated_values}
	\end{table}

	A similar, and exact, 	scaling can be derived for the case of photon scattering on a black disk, where the elastic differential cross-section is~\cite{Block:2006hy}

	\begin{align}
		\frac{d\sigma_{black}}{dt}=\pi R^{4}\left[\frac{J_{1}(q\cdot R)}{q\cdot R}\right]^{2}\,,
		\label{dssigmadt_black}
	\end{align}
	where $t=-q^{2}$ and $R$ is the radius of the black disk. The total cross-section is given by
	\begin{align}
		\sigma_{tot,black}=2\pi R^{2}\,.
		\label{sigmatot_black}
	\end{align}

	In this simple theoretical model the position of the first diffractive minimum, following from Eq.~(\ref{dssigmadt_black}), and the total cross-section Eq.~(\ref{sigmatot_black}) satisfies
	\begin{align}
		C_{black}=|t_{dip,black}|\cdot\sigma_{tot,black}=2\pi j_{1,1}^{2}(\hbar c)^2\approx~35.9\,\text{mb GeV}^{2}\,,
		\label{Cblack}   
	\end{align}
	where $j_{1,1}$ is the first root of the first order Bessel-function of the first kind~$J_{1}(x)$.

	The scaling behavior indicated by the stability of the value $C_{exp}$ is somewhat different from the black disk model, described
	by Eq.~(\ref{Cblack}), as the corresponding value $C_{black}$ is significantly different
	\begin{align}
		C_{black}\ne C_{exp}\,.
		\label{CblackCexp}
	\end{align}
	In this sense the value of $C_{exp}$ indicates a more complex scattering phenomena, than the photon black disc scattering.

\section{Summary and conclusions}

The real part of the forward scattering amplitude (FSA) is derived from unitarity constraints in the Bialas-Bzdak model leading to the so-called ReBB model. The added real part of the FSA significantly improves the model ability to
describe the data at the first diffractive minimum. In total the ReBB model describes both the ISR
and LHC data in the $0<|t|<2.5$~GeV$^{2}$ squared momentum transfer range in a statistically acceptable manner; in the latter case the fit range has to be divided to two parts, according to the
compilation of the two independent TOTEM measurements. The results are collected in Table~\ref{table:fit_parameters}.

The fit results also permit us to evaluate the shadow
profile functions $A(b)$, see Fig.~\ref{BBm_model_fit_results_shadow}. The plots indicate a Gaussian shape at ISR energies, while at LHC a saturation effect can be observed: the innermost part of the shadow profile function $A(b)$ around $b=0$
is almost flat and close to $A(b)\approx1$.
The elastic differential cross-section can be compared to a purely exponential distribution and the comparison shows a significant deviation from pure exponential in the $0.0\le|t|\le0.2$~GeV$^{2}$ range.

The fit results allow the determination of the excitation functions of the ReBB
model at future LHC energies and beyond, with parameters collected in Table 2 and
predicted differential cross-section curves shown in Fig.~\ref{BBm_model_extrapolation}. The shadow profile functions can
be also extrapolated, see Fig.~\ref{BBm_model_extrapolation_shadows}, which predicts that the saturated part of the proton
is expected to increase with increasing center of mass energy $\sqrt{s}$. The edge of the
distribution, the ``skin-width'' of the proton, expected to remain
approximately constant. It is worth to mention that the extrapolated version of the ReBB model utilizes of only eight parameters, the $p_{i}$ parameters of Table~\ref{extrapolation_parameters}, and
in this sense a ``minimal`` set of parameters is applied.

\section*{Acknowledgement}
	The authors are grateful to G.~Gustafson and L.~Jenkovszky for inspiring and fruitful discussions. This work was supported by the OTKA grant NK 101438 (Hungary) and the Ch.~Simonyi Fund (Hungary).


\begin{thebibliography}{99}


\bibitem{Bialas:2006qf}
  A.~Bialas and A.~Bzdak,
  Acta Phys.\ Polon.\ B {\bf 38} (2007) 159
  [hep-ph/0612038].

\bibitem{Nemes:2012cp}
  F.~Nemes and T. Cs\"org\H{o},
  Int.\ J.\ Mod.\ Phys.\ A {\bf 27} (2012) 1250175

\bibitem{CsorgO:2013kua}
  T. Cs\"org\H{o} and F.~Nemes,
  Int.\ J.\ Mod.\ Phys.\ A {\bf 29} (2014) 1450019


\bibitem{Ferreira:2014gda}
  E.~Ferreira, T.~Kodama and A.~K.~Kohara,
  arXiv:1411.3518 [hep-ph].

\bibitem{Kohara:2014waa}
  A.~K.~Kohara, E.~Ferreira and T.~Kodama,
  Eur.\ Phys.\ J.\ C {\bf 74} (2014) 11,  3175
  [arXiv:1408.1599 [hep-ph]].

\bibitem{Simone:WPCF2014}
S. Giani, "Overview of TOTEM results on 
total cross-section, elastic scattering and diffraction at LHC",
talk given at WPCF 2014, Gy\"ongy\"os, Hungary, August 2014,
{\tt https://indico.cern.ch/event/300974/session/2/contribution/34 }.

\bibitem{Jenkovszky:2014yea}
  L.~Jenkovszky and A.~Lengyel,
  arXiv:1410.4106 [hep-ph].

\bibitem{Antchev:2011vs}
  G.~Antchev {\it et al.}{TOTEM Collaboration},
Europhys.\ Lett.\  {\bf 96} (2011) 21002

\bibitem{Antchev:2013gaa}
  G.~Antchev {\it et al.}  [TOTEM Collaboration],
  Europhys.\ Lett.\  {\bf 101} (2013) 21002.

\bibitem{Antchev:2013iaa}
  G.~Antchev {\it et al.}  [TOTEM Collaboration],
  Europhys.\ Lett.\  {\bf 101} (2013) 21004.

\bibitem{Glauber_lectures}
R. J. Glauber, Lectures in Theoretical Physics, Vol. 1. Interscience, New York
1959.

\bibitem{Levin:1998pk}
  E.~Levin,
  hep-ph/9808486.

\bibitem{Khoze:2014aca}
  V.~A.~Khoze, A.~D.~Martin and M.~G.~Ryskin,
  arXiv:1402.2778 [hep-ph].

\bibitem{Ryskin:2012az}
  M.~G.~Ryskin, A.~D.~Martin and V.~A.~Khoze,
  Eur.\ Phys.\ J.\ C {\bf 72} (2012) 1937
  [arXiv:1201.6298 [hep-ph]].

\bibitem{Ryskin:2009qf}
  M.~G.~Ryskin, A.~D.~Martin, V.~A.~Khoze and A.~G.~Shuvaev,
  J.\ Phys.\ G {\bf 36} (2009) 093001
  [arXiv:0907.1374 [hep-ph]].


\bibitem{Martin:2012nm}
  A.~D.~Martin, H.~Hoeth, V.~A.~Khoze, F.~Krauss, M.~G.~Ryskin and K.~Zapp,
  PoS QNP {\bf 2012} (2012) 017

\bibitem{Lipari:2013kta} 
  P.~Lipari and M.~Lusignoli,
  Eur.\ Phys.\ J.\ C {\bf 73}, 2630 (2013)
  [arXiv:1305.7216 [hep-ph]].

\bibitem{Nagy:1978iw}
  E.~Nagy, R.~S.~Orr, W.~Schmidt-Parzefall, K.~Winter, A.~Brandt, F.~W.~Busser, G.~Flugge and F.~Niebergall {\it et al.},
      Nucl.\ Phys.\ B {\bf 150} (1979) 221.

\bibitem{Amaldi:1979kd}
  U.~Amaldi and K.~R.~Schubert,
      Nucl.\ Phys.\ B {\bf 166} (1980) 301.
\bibitem{Cheng:1969eh}
  H.~Cheng and T.~T.~Wu,
  Phys.\ Rev.\ Lett.\  {\bf 22} (1969) 666.

\bibitem{Cheng:1969bf}
  H.~Cheng and T.~T.~Wu,
  Phys.\ Rev.\  {\bf 182} (1969) 1852.

\bibitem{Cheng:1969ac}
  H.~Cheng and T.~T.~Wu,
  Phys.\ Rev.\  {\bf 182} (1969) 1868.

\bibitem{Chengbook}
H. Cheng and T.T. Wu, Expanding Protons: Scattering at High Energies , M.I.T. Press, Cambridge, MA (1987)

\bibitem{Bourrely:1978da}
  C.~Bourrely, J.~Soffer and T.~T.~Wu,
  Phys.\ Rev.\ D {\bf 19} (1979) 3249.

\bibitem{Bourrely:2014efa}
  C.~Bourrely, J.~Soffer and T.~T.~Wu,
  arXiv:1405.6698 [hep-ph].

\bibitem{Block:2006hy}
  M.~M.~Block,
  Phys.\ Rept.\  {\bf 436} (2006) 71
  [hep-ph/0606215].

\end{thebibliography}
\end{document}